%% file: MWCombinationEPJC.tex
\documentclass[epj]{svjour}

\usepackage{amssymb}

\usepackage{graphics}
\usepackage{graphicx}
\usepackage{subcaption}

\newcommand*{\Wlnu}{\ensuremath{W \rightarrow \ell\nu}}
\newcommand*{\met}{\ensuremath{\pT^{\nu}}}
\newcommand*{\ifb}{\mbox{fb\(^{-1}\)}}
\newcommand*{\pTl}{\ensuremath{\pT^{\ell}}}
\newcommand*{\pTnu}{\ensuremath{\pT^{\nu}}}

\newcommand*{\etal}{\ensuremath{\eta_{\ell}}}

\newcommand*{\mT}{\ensuremath{m_{\scriptsize\textrm{T}}}}
\newcommand*{\uT}{\ensuremath{u_{\scriptsize\textrm{T}}}}

\newcommand*{\pTW}{\ensuremath{\pT^{W}}}
\newcommand*{\pTZ}{\ensuremath{\pT^{Z}}}

\newcommand*{\pT}{\ensuremath{p_{\scriptsize\textrm{T}}}}

\newcommand*{\instlum}{\ensuremath{\textnormal{cm}^{-2}\textnormal{s}^{-1}}}

\begin{document}

\title{Compatibility and combination of world $W$-boson mass measurements}

\author{S.~Amoroso\inst{1} \and N.~Andari\inst{2} \and W.~Barter\inst{3} \and J.~Bendavid\inst{4} \and
  M.~Boonekamp\inst{2} \and S.~Farry\inst{5} \and M.~Gr{\"u}newald\inst{6} \and C.~Hays\inst{7} \and R.~Hunter\inst{8}
  \and J.~Kretzschmar\inst{5} \and O.~Lupton\inst{8} \and M.~Pili\inst{7} 
  \and  M.~Ramos Pernas\inst{8} \and B.~Tuchming\inst{2}
\and M.~Vesterinen\inst{8} \and A.~Vicini\inst{9} \and C.~Wang\inst{10} \and M.~Xu\inst{8} (LHC-TeV MW Working Group)}

\institute{DESY, Hamburg, Germany \and CEA/IRFU, Gif-sur-Yvette, France \and University of Edinburgh, Edinburgh, UK \and
  Massachusetts Institute of Technology, Cambridge, Massachusetts, USA \and University of Liverpool, Liverpool UK \and
  University College Dublin, Dublin, Ireland \and
  University of Oxford, Oxford, UK \and University of Warwick, Warwick, UK \and
  University of Milan, Milan, Italy \and Johannes Gutenberg University, Mainz, Germany}

\abstract{The compatibility of $W$-boson mass measurements performed by the ATLAS, LHCb, CDF, and D0 experiments is
  studied using a coherent framework with theory uncertainty correlations.
  The measurements are combined using a number of recent sets of parton distribution functions (PDF),
  and are further combined with the average
value of measurements from the Large Electron-Positron collider.  The considered PDF sets generally have a low
compatibility with a suite of global rapidity-sensitive Drell-Yan measurements.  The most compatible set is
CT18 due to its larger uncertainties.  A combination of all $m_W$ measurements yields a value of
$m_W = 80394.6 \pm 11.5$~MeV with the CT18 set, but has a probability of compatibility of 0.5\% and is
therefore disfavoured.  Combinations are performed removing each measurement individually, and a 91\%
probability of compatibility is obtained when the CDF measurement is removed.  The corresponding value of
the $W$ boson mass is $80369.2 \pm 13.3$~MeV, which differs by $3.6\sigma$ from the CDF value determined
using the same PDF set.
}

\maketitle

\section{Introduction}
\label{sec:intro}
\input{introduction}

\section{Overview of the measurements}
\label{sec:overview}
\input{overview}

\section{Methods}
\label{sec:methods}
\input{methods}

\subsection{Monte Carlo event generation}
\label{sec:generators}
\input{generators}

\subsection{Detector simulations}
\label{sec:simulation}
\input{simulation}

\section{$W$-boson production and decay}
\label{sec:production}
\input{production}

\subsection{$W$-boson \pT~distribution}
\label{sec:corrPTW}
\input{wpt}

\subsection{Parton distribution functions} 

\subsubsection{Central values and uncertainty correlations}
\label{sec:corrPDF}
\input{pdfextrap}

\subsubsection{$W$- and $Z$-boson production measurements}
\label{sec:PDFWZ}
\input{pdf}

\subsection{$W$-boson polarization}
\label{sec:polarization}
\input{polarization}

\subsection{$W$-boson resonance}
\label{sec:resonance}
\input{invmass}

\subsection{Electroweak corrections}
\label{sec:corrEWK}
\input{ewk}

\section{Combination}
\subsection{Procedures}
\label{sec:procedures}
\input{procedures}

\subsection{Results}
\label{sec:results}
\input{results}

\section{Conclusion}
\label{sec:conclusion}
\input{conclusion}

\section*{Acknowledgements}
We thank Joshua Isaacson, Pavel Nadolsky, Frank Tackmann, and C.-P. Yuan for fruitful discussions.
We are grateful to Joshua Isaacson and Yao Fu for providing us with the necessary inputs for generating events
with Resbos2.
We thank the ATLAS, CDF, D0 and LHCb collaborations for providing inputs to this analysis, and the CMS collaboration
for participating in the working group.
We thank
Raymond Brock, Paul Grannis, Rick Van Kooten, Hugh Montgomery, Pierre Petroff, Heidi Schellman,
Bill Ashmanskas, Bo Jayatilaka, Mark Lancaster,
Stefano Camarda, Aleksandra Dimitrievska, Monica Dunford, Pamela Ferrari, Oldrich Kepka, Bogdan Malaescu, Philip Sommer,
Guillaume Unal,
Hengne Li, Katharina Mueller, Matthew Needham, Monica Pepe-Altarelli, Federico Redi, Lorenzo Sestini, Frederic Teubert, and
Hang Yin for useful comments and suggestions.
We thank the LHC Electroweak Working Group for facilitating useful discussions and providing computing
resources for this work.
MV is supported by the ERC-CoG-865469 SPEAR grant, WB is supported by the UKRI Future Leaders Fellowships grant MR/W009048/1,
and SA is supported by the Helmholtz Association contract W2/W3-123.

 \appendix

 \section{Further information}
\label{app:correlations}
\input{appendix}

 \bibliographystyle{elsarticle-num}
  \bibliography{MWCombinationEPJC.bib}

\end{document}

%% file: introduction.tex
The $W$-boson mass ($m_W$) is an important parameter of the Standard Model (SM) of particle physics, providing a 
sensitive test of the model's consistency and offering a window to potential new processes.  An active program of
measurements at the Tevatron and Large Hadron Collider (LHC) continues to improve the experimental
precision of $m_W$, which is approaching the uncertainty on the SM prediction.  Previous measurements from the Large
Electron Positron collider (LEP) together have a precision comparable to the individual hadron-collider measurements.
A combination of the Tevatron, LHC, and LEP measurements can thus improve the precision on $m_W$ and quantify the
compatibility of the measurements.  Such a compatibility study is particularly motivated in light of the discrepancy
between the most recent measurement~\cite{CDF:2022hxs} from the CDF experiment at the Tevatron and previous
measurements~\cite{Abazov:2009cp,D0:2013jba,LHCb:2021bjt,Aaboud:2017svj} from the D0 experiment at the Tevatron,
and the LHCb and ATLAS experiments at the LHC. 

At hadron colliders, measurements of $m_W$ exploit the kinematic peaks of distributions observed in leptonic 
$W$-boson decays.  These final-state distributions carry information about the decaying particle mass, but also
depend on other $W$-boson degrees of freedom such as the $W$-boson rapidity, transverse momentum, and polarization.  
Predictions of these distributions are generally obtained using Monte Carlo (MC) event generators with input parton 
distribution functions (PDF).  Past measurements have used different generators and PDF sets, so prior to combining
the measurements a coherent treatment is required to compare measurements and obtain uncertainty correlations.  Where
appropriate, small adjustments are thus applied to the measured values or uncertainties.  These adjustments are
estimated using a fast detector simulation developed for this purpose, or using the simulation from the experimental
measurement.

The presentation of the combination begins with an overview of the individual measurements in Sec.~\ref{sec:overview}, 
followed by a description of the methods in Sec.~\ref{sec:methods}.  The theoretical treatment of the $W$-boson 
production and decay is provided in Sec.~\ref{sec:production}, along with uncertainties, correlations, and any 
adjustments to the measurements.  The results of the combination are presented in Sec.~\ref{sec:results}, and 
conclusions are given in Sec.~\ref{sec:conclusion}.

%% file: overview.tex
The combination uses the latest measurements from D0 and CDF at the Tevatron and ATLAS and LHCb at the LHC.
The CMS Collaboration has not yet measured $m_W$, though it has measured differential $W$-boson cross sections
on the path to the measurement~\cite{CMSW}.
Prior measurements from the Tevatron and the CERN Super Proton Synchroton are not included as they are expected
to have negligible impact.  
The hadron-collider measurements are combined with
the result from the Large Electron Positron collider (LEP)~\cite{ALEPH:2013dgf},
$m_W = 80.376 \pm 0.033$~GeV\footnote{We use the convention $c \equiv 1$ and work in a right-handed coordinate system
with the origin at the centre of the detector and the $z$-axis along the beam pipe.  Cylindrical coordinates $(r,\phi)$
are used in the transverse plane, where $\phi$ is the azimuthal angle around the z-axis.  The pseudorapidity is defined
in terms of the polar angle $\theta$ as $\eta = -\ln\tan(\theta/2)$.  Transverse momentum is defined as $\pT = p\sin\theta$.
}.

The kinematic observables used in $m_W$ measurements at hadron colliders are the momentum of the charged lepton from
the $W$-boson decay ($p^{\ell}_{\scriptsize\textrm{T}}$) and the recoil transverse momentum ($u_{\scriptsize\mathrm{T}}$)
balancing the transverse momentum of the $W$ boson (\pTW).  The recoil is measured by vectorially summing the momentum of
all objects interacting in the detector, except for the charged lepton.  The neutrino momentum is inferred from the
net momentum imbalance,
$\vec{p}^{\nu}_{\scriptsize\mathrm{T}} \equiv -(\vec{p}^{\ell}_{\scriptsize\mathrm{T}} + \vec{u}_{\scriptsize\mathrm{T}})$.
For experiments with sufficiently good recoil resolution the most sensitive kinematic distribution is the transverse mass,
$m_{\scriptsize\textrm{T}} = \sqrt{2p_{\scriptsize\textrm{T}}^{\ell} p_{\scriptsize\textrm{T}}^\nu (1-\cos\Delta\phi)}$,
where $\Delta \phi$ is the angle between the charged lepton and the neutrino in the transverse plane.  

The CDF Collaboration measured $m_W$~\cite{CDF:2022hxs} using Run 2 data collected between 2003 and 2011 at the Tevatron
collider, corresponding to 8.8~\ifb{} of integrated luminosity from proton-antiproton ($p\bar{p}$) collisions at a
center-of-mass energy of $\sqrt{s}=1.96$~TeV.  The mass was obtained from template fits to the reconstructed
distributions of $p^{\ell}_{\scriptsize\mathrm{T}}$, \mT, and $p^{\nu}_{\mathrm T}$ in the 
electron and muon decay channels, yielding $m_W=80433.5\pm 6.4~\textrm{(stat.)} \pm 6.9~\textrm{(sys.)}$~MeV, or
$80433.5\pm 9.4$~MeV.  
The quoted value of $m_W$ corresponds to the NNPDF3.1 PDF set~\cite{Ball:2017nwa}, with the PDF uncertainty estimated using 
the largest 25 symmetric eigenvectors constructed through a principal-component analysis from the full replica
set.  The direct fit for $m_W$ to the data used events from a version of the {\textsc{ResBos}}~\cite{Ladinsky:1993zn}
generator referred to as {\textsc{ResBos-C}} in this paper.  The generation used the CTEQ6M PDF set~\cite{Pumplin:2002vw} 
and was tuned to fit the observed spectrum of $Z$-boson transverse momentum.  The uncertainty on the $W$-boson transverse 
momentum $p_{\scriptsize\textrm{T}}^{W}$ was determined using DYqT~\cite{Bozzi:2008bb,Bozzi:2010xn}, with a constraint from
the observed recoil
distribution in $W$-boson events.  The adjustment of the model to the NNPDF3.1 PDF set included an effective update of
the modelling of the leptonic angular distributions, as discussed in Sec.~\ref{sec:polarization}.

The D0 Collaboration performed two measurements of $m_W$ in Run 2 of the Tevatron collider.  The first used data taken
between 2002 and 2006, corresponding to an integrated luminosity of 1.1~\ifb{}~\cite{Abazov:2009cp}, and the second
used 2006-2008 data corresponding to an integrated luminosity of 4.3~\ifb{}~\cite{D0:2013jba}.  The analysis 
produced template fits for $m_W$ using the $p^{\ell}_{\scriptsize\textrm{T}}$, \mT, and $p^{\nu}_{\scriptsize\textrm{T}}$
kinematic distributions in the electron decay channel.  The initial 1.1~\ifb{} measurement combined the results from these
three distributions, while the 4.3~\ifb{} measurement removed the $p^{\nu}_{\scriptsize\textrm{T}}$ result due to its small
weight in the combination.  The overall combined result of all measurements is
$m_W = 80375\pm13~\textrm{(stat.)}\pm22~\textrm{(sys.)}$~MeV, or $80375\pm23$~MeV.  This value was determined using the
CTEQ6.1~\cite{Stump:2003yu} (CTEQ6.6~\cite{Nadolsky:2008zw}) PDF set for the measurement using 1.1~\ifb{} (4.3~\ifb{}).
The uncertainties were evaluated using {\textsc{Pythia6}}~\cite{Sjostrand:2006za} and the CTEQ6.1 PDF Hessian eigenvectors
scaled to reduce the nominal 90\% C.L. coverage to 68\% C.L.  The $p_{\scriptsize\textrm{T}}^{W}$ modelling used a version of
the {\textsc{ResBos}}~\cite{Balazs:1997xd,Landry:2002ix} generator referred to here as {\textsc{ResBos-CP}}.  

The $m_W$ measurement performed by the ATLAS Collaboration used $\sqrt{s}=7$~TeV proton-proton collision data
corresponding to 4.6~\ifb{} of integrated luminosity collected in 2011 during Run 1 of the LHC collider.
ATLAS performed template fits to the $p^{\ell}_{\scriptsize\textrm{T}}$ and \mT~distributions in the electron and muon
channels separately for
$W^+$ and $W^-$ events, since in proton-proton ($pp$) collisions the final-state distributions are different for these
processes.  The fits were further subdivided into three (four) pseudorapidity ranges in the electron (muon) channel,
yielding a total of 28 measurements.  The combination of these measurements yields 
$m_W=80370\pm7~\textrm{(stat.)}\pm18~\textrm{(sys.)}$, or $80370\pm19$~MeV.  The parton distribution functions were
modelled with the NNLO CT10 PDF set~\cite{Lai:2010vv}, with the Hessian uncertainties scaled to 68\% C.L.  The
$p_{\scriptsize\textrm{T}}^{W}$ modelling relied on the parton shower Monte Carlo (MC)
{\textsc{Pythia8}}~\cite{Sjostrand:2014zea} tuned to match the $p_{\scriptsize\textrm{T}}^{Z}$ distribution observed in
data.  The impact of the PDF uncertainties on the $m_W$ measurement was reduced by a simultaneous fit in different lepton
pseudorapidity regions.  The PDFs affect both the $p_{\scriptsize\textrm{T}}^{W}$ and $p_{\scriptsize\textrm{T}}^{Z}$
distributions, and to preserve the agreement with the $p_{\scriptsize\textrm{T}}^{Z}$ data distribution only the relative
variation between the $p_{\scriptsize\textrm{T}}^{W}$ and $p_{\scriptsize\textrm{T}}^{Z}$ distributions was propagated in
the uncertainty estimate.  Generated events were reweighted according to the calculation of the leptonic angular 
distributions in DYNNLO~\cite{Catani:2007vq,Catani:2009sm}.

The LHCb Collaboration performed a measurement of $m_W$ using Run 2 $pp$ LHC collision data collected in 2016 at 
$\sqrt{s}=13$~TeV, corresponding to 1.7~\ifb{} of integrated luminosity.  The measurement used the
$q/p^{\ell}_{\scriptsize\textrm{T}}$ distribution in the muon decay channels, where $q$ is the muon charge, giving a result
of $m_W = 80354 \pm 23~\textrm{(stat.)} \pm 10~\textrm{(exp.)} \pm 17~\textrm{(th.)}\pm 9~\textrm{(PDF)}$~MeV, or
$80354\pm32$~MeV.  The LHCb central value of $m_W$ and its uncertainty correspond to an unweighted average of results
using the
NNPDF3.1, MSHT2020~\cite{Bailey:2020ooq} and CT18~\cite{Hou:2019qau} PDF sets, all at next-to-leading order in the strong
coupling and with 68\% C.L. coverage.  The \pTW~distribution was modelled with
{\textsc{Powheg}~\cite{Nason:2004rx,Frixione:2007vw,Alioli:2010xd} interfaced to {\textsc{Pythia8}}, with a correction
at high boson $p_{\scriptsize\textrm{T}}$ derived from the observed $Z$-boson $p_{\scriptsize\textrm{T}}$ distribution.
The leptonic angular distributions were modelled with exact
$\mathcal{O}(\alpha_{\scriptsize\textrm{S}}^2)$ predictions from DYTurbo~\cite{Camarda:2019zyx}
and modified by scaling one of the leptonic angular coefficients when fitting the data.

The event requirements and fit ranges used in the measurements are summarized in Table~\ref{tab:evsel}.
CDF and D0 used similar analysis configurations, while at
ATLAS the looser recoil requirement and wider \mT~fit range were a consequence of the lower recoil resolution.
The LHCb measurement was inclusive in recoil, with only a loose requirement on the momentum of the muon.
The ATLAS, CDF, and D0 measurements fit $m_W$ only, while LHCb performed a simultaneous fit for $m_W$ and the
relative fraction of $W^+$- to $W^-$-boson decays, the hadronic background fraction, $\alpha_{\scriptsize\textrm{S}}$
in $W$-boson events, $\alpha_{\scriptsize\textrm{S}}$ in $Z$-boson events, the intrinsic transverse momentum 
distribution of partons inside the proton, and the $A_3$ leptonic angular coefficient (see Sec.~\ref{sec:polarization}).

\begin{table}[!t]
  \centering
  \begin{tabular}{lll}
\hline
\hline
    Experiment & Event requirements & Fit ranges \\
\hline
    \vspace{-3mm}\\
    CDF   & $30<\pTl<55$~GeV & $32<\pTl<48$~GeV \\
          & $|\etal|<1$      & $32<\met<48$~GeV \\
          & $30<\met<55$~GeV & $60<\mT<100$~GeV \\
          & $65<\mT<90$~GeV  & \\
          & $\uT<15$~GeV  & \\
\vspace{-3mm}\\
    D0    & $\pT^e >25$~GeV  & $32<\pT^e<48$~GeV \\
          & $|\etal|<1.05$   & $65<\mT<90$~GeV  \\
          & $\met>25$~GeV    & \\
          & $\mT>50$~GeV     & \\ 
          & $\uT<15$~GeV     & \\
    \vspace{-3mm}\\
    ATLAS & $\pTl>30$~GeV  & $32<\pTl<45$~GeV \\
          & $|\etal|<2.4$  & $66<\mT<99$~GeV  \\
          & $\met>30$~GeV  & \\
          & $\mT>60$~GeV   & \\
          & $\uT<30$~GeV   & \\
    \vspace{-3mm}\\
    LHCb  & $\pT^\mu > 24$~GeV & $28 < \pT^\mu < 52$~GeV \\ 
          & $2.2 < \eta_\mu < 4.4$ & \\  
    \vspace{-3mm}\\
\hline \hline
  \end{tabular}
  \caption{Event requirements and fit ranges for CDF, D0, ATLAS, and LHCb.\label{tab:evsel}}
\end{table}

%% file: methods.tex
The combination consists of three steps.  First, the results are adjusted to a common model to allow 
a consistent comparison of central values and evaluation of uncertainty correlations.  This reference
model includes the description of the $W$-boson production, the Breit-Wigner lineshape, and the $W$-boson
polarization, and is described in Section~\ref{sec:production}.  Second, the correlation of
uncertainties between the experiments is evaluated.  The different center-of-mass energies, initial states,
and lepton pseudorapidity coverage make the correlation non-trivial.
Finally, the results are combined for representative PDF sets, with the compatibility of the measurements
determined for each set.  In addition, other $W$ and $Z$ boson measurements at the Tevatron and LHC are
compared to predictions using these PDF sets, in order to study the reliability of the PDF predictions and
uncertainties for the $m_W$ measurement.

The first step of adjusting each result to a different theoretical model requires an emulation of the 
measurement process, which consists of Monte Carlo event generation (see Sec.~\ref{sec:generators}), detector
simulation (see Sec.~\ref{sec:simulation}), event selection, and a kinematic fit for $m_W$.  The Monte Carlo
samples are produced using a reference value $m_W^{\scriptsize\textrm{ref}}$ for the $W$-boson mass and width
(${\mathrm \Gamma}_W$), and different values of $m_W$ are obtained by reweighting events according to a Breit-Wigner
distribution,
\begin{equation}
  w(m,m_W,m_W^{\scriptsize\textrm{ref}}) = \frac{(m^2 - m_W^2)^2 + m^4 {\mathrm \Gamma}_W^2 / m_W^2} 
{(m^2 - {m_W^{\scriptsize\textrm{ref}}}^2)^2 + m^4 {\mathrm \Gamma}_W^2 /   {m_W^{\scriptsize\textrm{ref}}}^2},
\end{equation}
using the final-state invariant mass $m$.  
This parameterization uses the running width scheme in accordance with 
the published measurement procedures.  The mass reweighting procedure has been checked to give the 
correct target mass value within a statistical uncertainty of $\approx 0.2$~MeV.

The detector simulations used in the original ATLAS, CDF, and D0 measurements are simplified so that
large event samples can be simulated for a variety of PDF sets (see Section~\ref{sec:simulation}).
These simulations do not have the complexity required for a mass measurement in data but are sufficient
for estimating the impact of small theoretical modifications on the measurement.  For the LHCb measurement
no simplification is necessary and the same detector simulation is used as in the original measurement.

The shift in the value of $m_W$ resulting from a change in the generator model is estimated by producing 
template distributions using a given experiment's model, and the same kinematic distributions for an 
alternate model (the “pseudo-data”).  The shift is determined by minimizing the negative log-likelihood
between the pseudo-data and the template distributions.
In the following we quote the impact $\delta m_W$ of each theoretical shift on a
measurement, i.e. the change in the measured $m_W$ value for a given change in the theoretical model.

A common set of uncertainties and correlations between experiments is obtained by evaluating $\delta m_W$ 
for a variety of PDF sets within the reference theoretical model.  Summing the theoretical shifts gives a
total $\delta m_W$ that we add to each experimental measurement to obtain the value to be used in the
combination.  For each PDF set the combination is performed using the method 
of the best linear unbiased estimator (BLUE)~\cite{Valassi:2003mu}, including both theoretical and experimental 
uncertainties.  The BLUE method is used by the individual experiments to combine results from different 
kinematic distributions, and the combination procedure is validated by reproducing each experiment's published 
value.  Results are presented for a combination of all experimental measurements, as well as for various 
measurement subsets.

%% file: generators.tex
The effects of modifying the $W$-boson production and decay model are studied using 
event samples for the \Wlnu~process in $pp$ collisions at $\sqrt{s}=7$~TeV and
$\sqrt{s}=13$~TeV, and for $p\bar{p}$ collisions at $\sqrt{s}=1.96$~TeV.
The event generators include those used by the original experiments, along with more
recent versions with improved calculations.  The PDF sets considered include those
from the original measurements (CTEQ6M, CTEQ6.1, CTEQ6.6, CT10, NNPDF3.1, CT18,
and MSHT2020), as well as the following sets at next-to-next-to-leading order (NNLO) in
$\alpha_{\scriptsize\textrm{S}}$: NNPDF4.0~\cite{NNPDF:2021njg},
ABMP16~\cite{Alekhin:2017kpj}, CT14~\cite{Dulat:2015mca}, and MMHT2014 \cite{Harland-Lang:2014zoa}.

For the ATLAS, CDF, and D0 experiments the $m_W$ shift associated with a particular NNLO PDF set is
evaluated using the NNLO QCD calculation {\textsc{Wj-MiNNLO}} \cite{Monni:2019whf,Monni:2020nks}
implemented in {\textsc{Powheg-Box-V2}}~\cite{Nason:2004rx,Frixione:2007vw,Alioli:2010xd}.  The analysis 
is performed at the Les Houches event level~\cite{Alwall2007} without interfacing to a parton shower,
allowing for an efficient and fast processing.  The addition of the parton shower has been confirmed
to negligibly affect the shift associated with the PDF set.  A large sample is produced for each PDF set,
with weights calculated internally by \textsc{Wj-MiNNLO} in order to evaluate PDF uncertainties. 

The uncertainty associated with a given PDF set is evaluated using the next-to-leading order (NLO) QCD
calculation {\textsc{W\_ew-BMNNP}}~\cite{Barze:2012tt} implemented in the {\textsc{Powheg-Box-V2}}.  
This calculation is used for efficiency reasons and the difference in the estimated uncertainty
with respect to an NNLO calculation is expected to be negligible.
For LHCb the $m_W$ shifts and uncertainties are evaluated using
the same {\textsc{Powheg}} NLO calculation of $W$-boson production~\cite{Alioli:2008gx} as used in the original
measurement.

The modelling of the $W$-boson polarization and resonance lineshape are studied using
large \textsc{ResBos} samples corresponding to those from the Tevatron measurements:
{\textsc{ResBos-C}}~\cite{Balazs:1997xd}, used by CDF in their direct fit to the
data~\cite{CDF:2022hxs}, with an accuracy of NLO and approximate NNLL in
QCD~\cite{Isaacson:2022rts}; and {\textsc{ResBos-CP}}~\cite{Landry:2002ix}, used in the
D0 $m_W$ measurement~\cite{D0:2013jba}, with an accuracy of NNLO+NNLL in QCD.  A third sample is 
generated using {\textsc{ResBos2}}~\cite{Isaacson:2022rts} with an accuracy of NLO+NNLL in QCD and
including a full resummation of the coefficients describing the leptonic angular distributions (see
Section~\ref{sec:polarization}).  The difference between NLO and NNLO predictions of these
coefficients is studied using the DYNNLO generator~\cite{Catani:2007vq,Catani:2009sm}, which has
been confirmed to be consistent with other fixed-order calculations~\cite{comparisons}.

Electroweak corrections, primarily photon radiation in $W$-boson decay, have a large impact on the 
final-state distributions but are calculated accurately.  The experiments factorize these corrections
from PDF and QCD effects and we therefore do not include them in the sample generation.

%% file: simulation.tex
This subsection provides brief descriptions of the parameterized simulations used to study 
the effects of model variations on the combination, and shows fit distributions comparing the 
simulations to those used in the experiments.  The simulation of each detector is referred to 
as the ``LHC-TeV MWWG'' or ``MWWG'' simulation for that detector.  A systematic uncertainty due to
the simplified parameterization is conservatively estimated by varying the lepton and recoil scales 
and resolutions by $\pm 5$\% and determining the effect on $\delta m_W$.

\subsubsection{CDF response and resolution model}

The CDF detector model consists of parameterizations of the electron and recoil momentum response 
and resolution.  The muon momentum response and resolution are similar to those of electrons. 

The electron fractional momentum resolution $\sigma_{p_{\tiny\textrm{T}}}/\pT = \sqrt{\kappa^2 + S^2/\pT}$ is modelled using
a sampling term of $S=12.6\%$~GeV$^{1/2}$ and a constant term of $\kappa=2\%$.  The constant term 
is larger than that used in the CDF simulation in order to correct for the lack of final-state radiation 
in the generated samples.  The leakage of the shower beyond the electromagnetic calorimeter is parameterized 
in the same manner as for the CDF measurement, and the reduction in electron momentum is corrected with a
scale factor applied to the electron momentum.

The recoil response is defined as the ratio $R(\pTW)$ of the measured recoil $u_T$ to the generated $\pTW$,
before accounting for effects of underlying event and additional proton-antiproton interactions (pileup).   
The CDF response is parameterized as 
\begin{equation}
R(\pTW) = r \log(a \pTW)/ \log(a \pT^{\scriptsize\textrm{ref}}),
\end{equation}

\noindent
where 
$r=0.65$, $a=6.7/$GeV, and $\pT^{\scriptsize\textrm{ref}} = 15$~GeV.  

The jet-like sampling for the recoil resolution is 
\begin{equation}
\sigma(\pTW) = s \sqrt{\pTW}, 
\end{equation}

\noindent
with 
$s=0.87$~GeV$^{1/2}$ and \pTW~in GeV. 
The recoil azimuthal angular resolution $\sigma_{u_\phi}$ is parameterized as 
\begin{equation}
\sigma_{u_\phi}(\pTW)  = \alpha - \beta \times \pTW,
\end{equation}

\noindent
where 
$\alpha=0.273$~rad and $\beta=0.016$~rad/GeV for $\pTW<\pT^{\scriptsize\textrm{ref}}$, and 
$\alpha=0.143$~rad and $\beta=0.0044$~rad/GeV for $\pTW \geq \pT^{\scriptsize\textrm{ref}}$.

The contribution of the underlying event to the measured recoil is represented by a randomly-oriented Gaussian 
distribution of width 
6.2~GeV.
Finally, the removal of lepton calorimeter towers from the recoil reconstruction is modelled by subtracting 660
MeV from the generated recoil along the direction of the decay lepton.

\begin{figure}[!tbp]
   \centering
   \includegraphics[width=0.98\columnwidth]{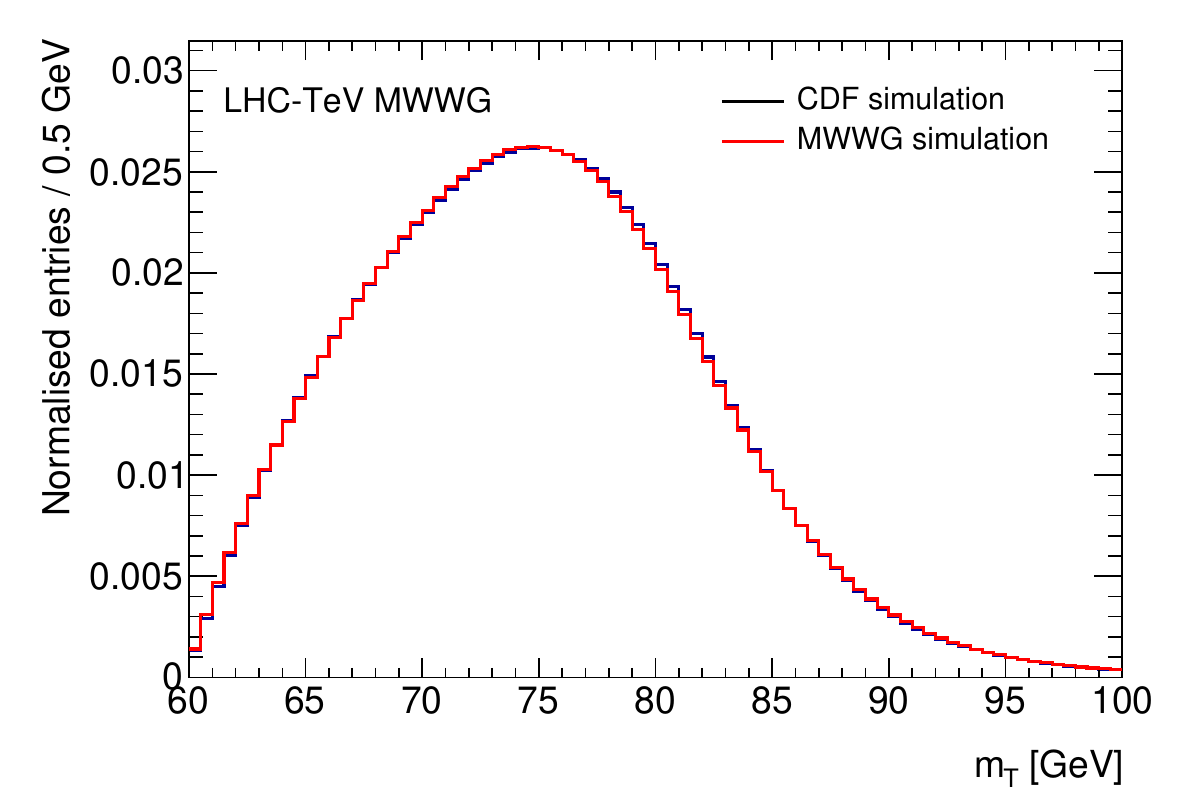}
   \includegraphics[width=0.98\columnwidth]{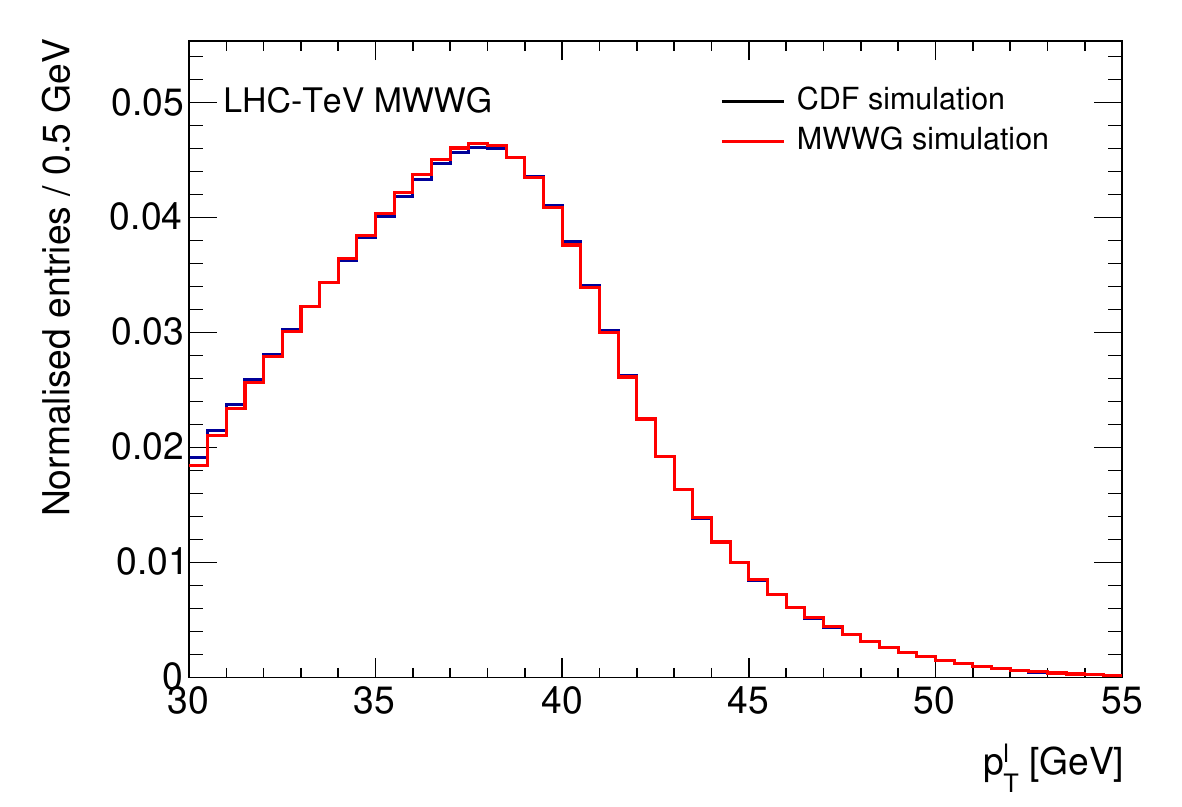}
   \caption{
Comparisons between the CDF simulation~\cite{Aaltonen:2013vwa} and the LHC-TeV MWWG simulation for 
the \mT~(top) and \pTl~(bottom) distributions.  
}
\label{fig:simuCDF}
\end{figure}

The distributions obtained using the MWWG simulation are compared with those from the CDF simulation 
in Fig.~\ref{fig:simuCDF}.  The agreement is at the percent 
level in the range of interest for the measurement.
The systematic uncertainties on the MWWG simulation are estimated by varying the response and resolution by $\pm 5\%$, 
calculating the $\delta m_W$ shifts for fourteen PDFs, and taking the maximum $|\delta m_W|$.  The resulting
uncertainties are 1.0 MeV for the \mT~fit, 0.9 MeV for the \pTl~fit, and 2.0 MeV for the \met~fit.
 
\subsubsection{D0 response and resolution model}

The LHC-TeV MWWG simulation of the D0 detector includes a model of the efficiency of the electron reconstruction and 
selection along with the response and resolution of the recoil and the electron momentum.  The simulation reproduces 
distributions from the D0 parameterized Monte Carlo simulation (PMCS) used for the final D0 measurement based on an 
integrated luminosity of 4.3~\ifb{}~\cite{D0:2013jba}.  The prior measurement based on 1.1~\ifb{}~\cite{Abazov:2009cp} 
of integrated luminosity has a lower mean number of pileup events; the corresponding impact on the estimation of
mass shifts is within the applied uncertainty.

The electron energy response is parameterised as:
\begin{equation}
E = \alpha(E_0-\bar{E_0}) + \beta + \bar{E_0},
\end{equation}
where $E$ is the calibrated electron energy, $\bar{E_0}=43$~GeV is a reference value corresponding to the electron energy 
in $Z$-boson events, and $\alpha$ and $\beta$ are luminosity-dependent energy scale and offset corrections, respectively.  
We take $\alpha=1.0164$ and $\beta=0.188$~GeV, the values determined in Ref.~\cite{D0:2013jba} for an instantaneous 
luminosity in the range (2--4)$\times 36\times 10^{30}$\instlum~corresponding to the largest fraction of the 
data.  Implementing the instantaneous luminosity dependence gives results in agreement with the average response to 
within a percent.

The electron energy resolution $\sigma_E/E$ is simulated using the same functional form as for CDF, with a constant term
of $\kappa=1.997\%$~\cite{D0:2013jba} and a sampling term of
\begin{eqnarray}
  S &=& S_0 \exp\left[S_1\left( \frac {1}{\sin\theta}-1\right)\right] + \frac{S_2\eta+S_3}{\sqrt{E}},
 \end{eqnarray}
where $S_0=0.153$~GeV$^{1/2}$, $S_1=1.543$, $S_2=-0.025$~GeV, $S_3=0.172$~GeV, and $E$ is in GeV. The resulting fractional
resolution is increased by 2\% to account for the lack of generated final-state radiation and improve the agreement with the
distributions from the D0 PMCS.

The electron reconstruction and identification efficiency is modeled by the following function
determined using the data points in Fig.~25(b) of Ref.~\cite{D0:2013jba}:
\begin{eqnarray}
\varepsilon( \pTl)&=& 0.95 \left(1- e^{-{0.074  \pTl}}\right),
\end{eqnarray}
where \pTl~is in GeV.

The recoil is modelled using a migration matrix to obtain a simulated \uT~value for a given generated
\pTW~\cite{D0:2020ujb}.  In order to model the recoil energy in the electron cone that is not included
in the recoil measurement, 150~MeV are subtracted from the recoil component parallel to the decay
lepton~\cite{D0:2013jba}.

Figure~\ref{fig:simuD0} shows the \pTl~and \mT~distributions from the D0 PMCS and the LHC-TeV MWWG 
simulation after reweighting the events to match the \pTW~distribution used for the D0 measurement.  The distributions
agree to within 2\% in the range of interest for the $m_W$ extraction.  The shifts in $m_W$ are studied for the 
eigenvectors of the CTEQ6.6 and CT10 PDFs, and the MWWG simulation and D0 PMCS agree within the statistical 
precision of $\approx 1$~MeV.  Systematic uncertainties are determined by varying the scales and resolutions, and 
calculating the effect on $\delta m_W$ for fourteen PDF sets.  The resulting uncertainties are 1.0 MeV on the \mT~fit, 
1.0 MeV on the \pTl~fits, and 2.2 MeV on the \met~fit.

\begin{figure}[!tbp]
  \centering
  \includegraphics[width=0.98\columnwidth]{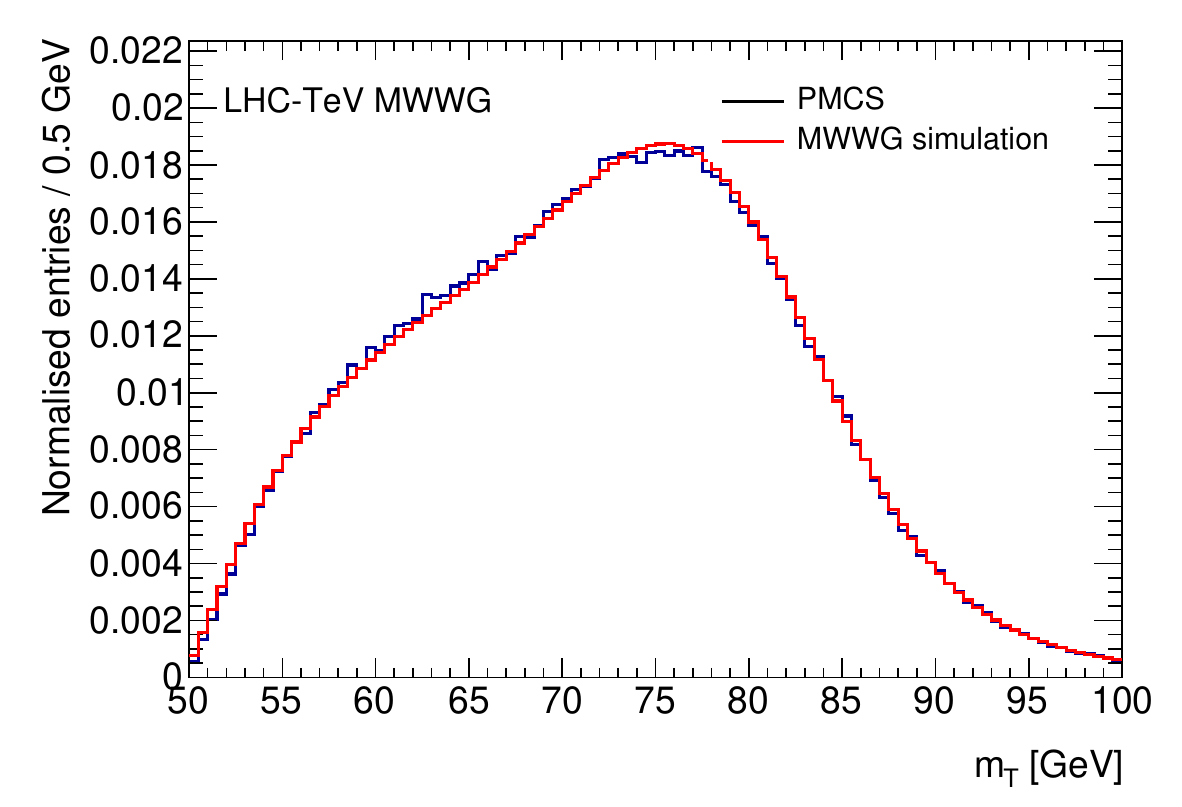}
  \includegraphics[width=0.98\columnwidth]{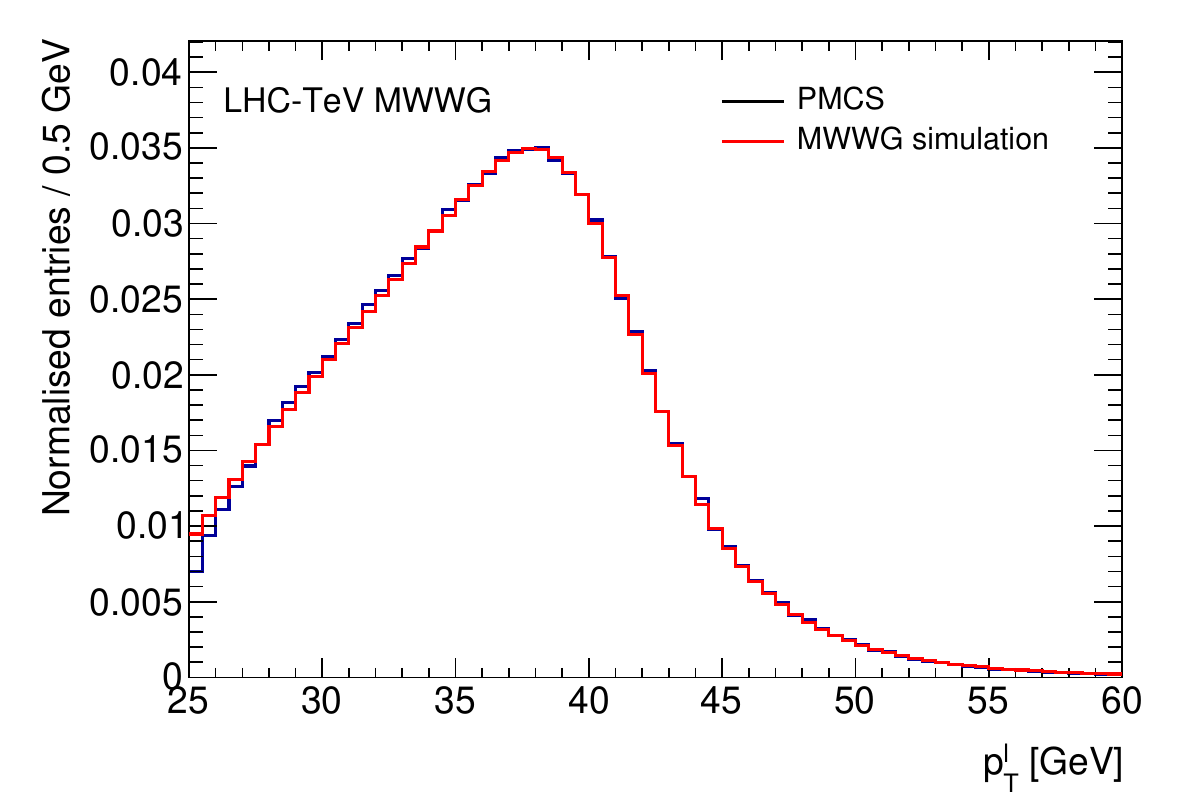}
  \caption{Comparisons of the D0 PMCS and the LHC-TeV MWWG simulation for the \mT~(top) and \pTl (bottom) distributions. 
}
\label{fig:simuD0}
\end{figure}

\subsubsection{ATLAS response and resolution model}

The ATLAS recoil response and resolution are parametrized using distributions~\cite{Aaboud:2017svj} of the projections
of these quantities
along and perpendicular to the lepton direction, as a function of the $W$-boson transverse momentum.  The parameterizations
are calibrated using the full ATLAS simulation.  The recoil resolution ranges from 12 to 16~GeV, depending primarily
on the amount of pileup.  
The electron and muon resolutions are parameterized using the documented detector performance~\cite{PERF-2013-05,PERF-2014-05}.
The resulting \pTl~and \mT~distributions are given in Figure~\ref{fig:simuATLAS}, which shows that the resolution is
accurately modeled and that residual differences could be improved with lepton energy scale adjustments and do not
significantly affect the results.

\begin{figure}[!tbp]
  \centering
  \includegraphics[width=0.98\columnwidth]{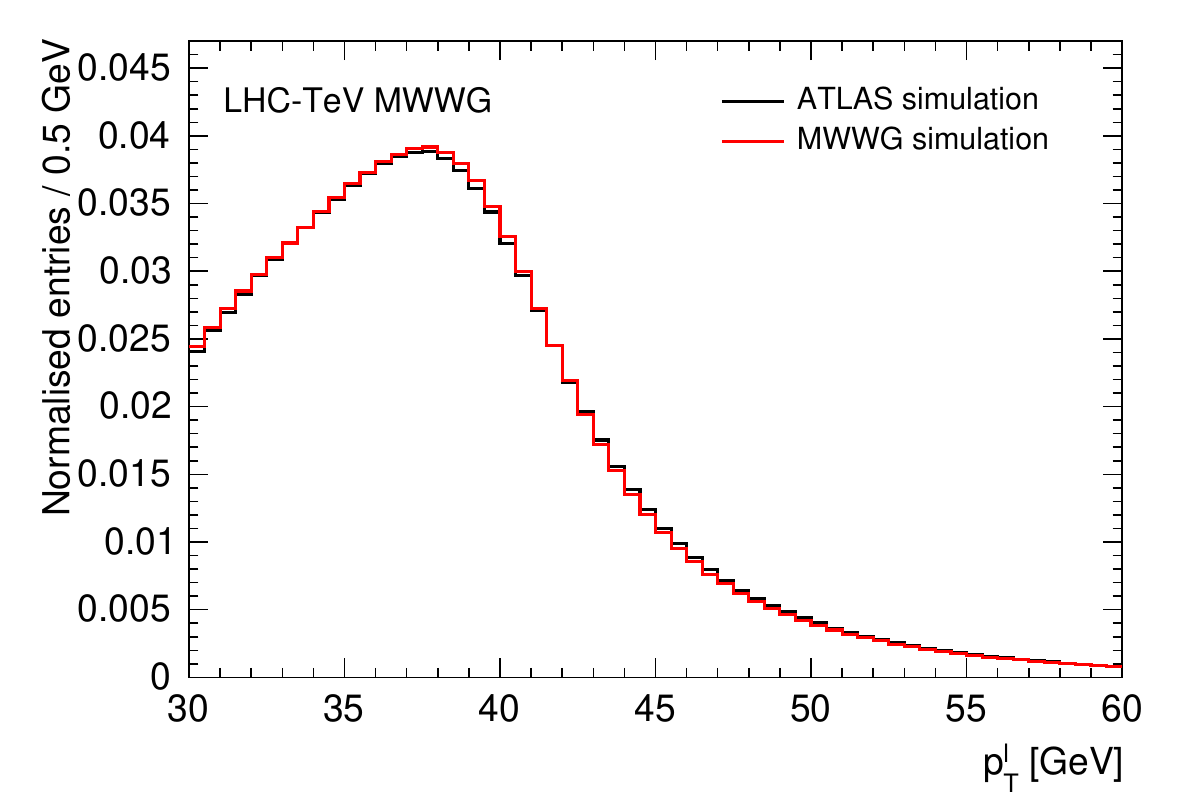}
  \includegraphics[width=0.98\columnwidth]{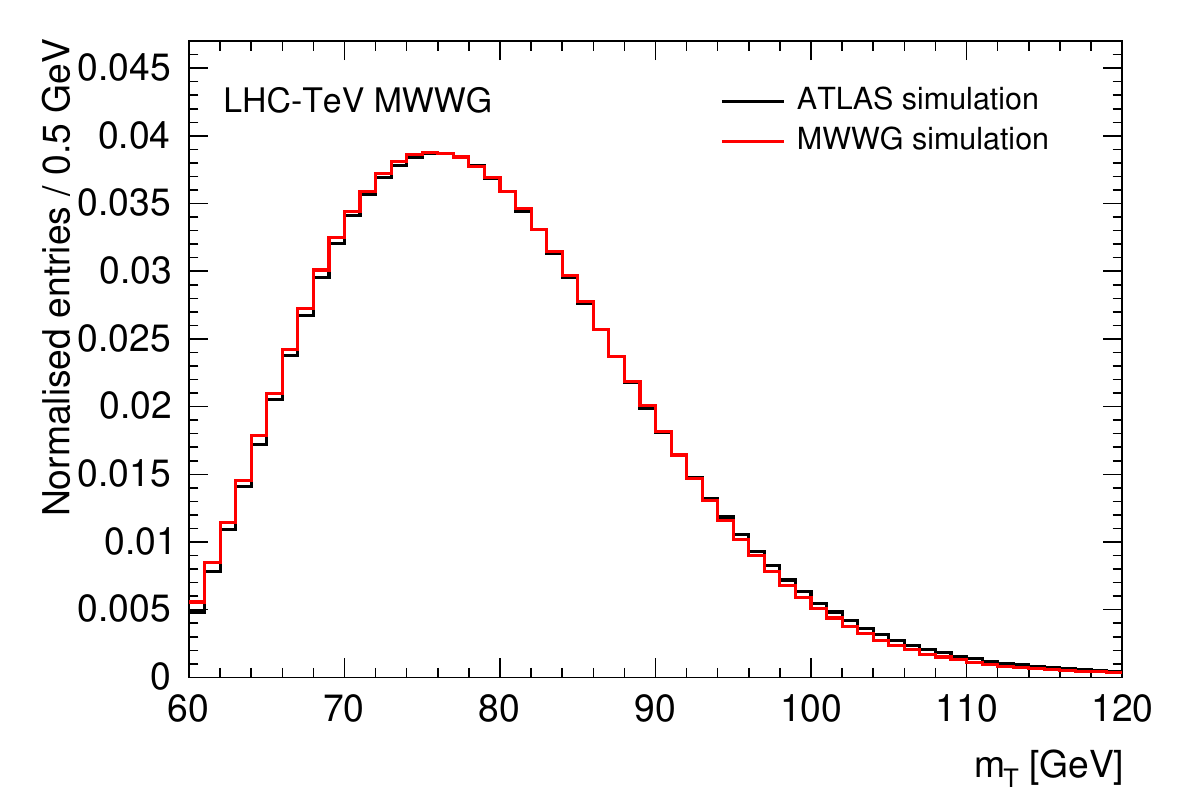}
  \caption{Comparison of the published and MWWG simulated \pTl~(top) and \mT~(bottom) distributions for ATLAS.
}
\label{fig:simuATLAS}
\end{figure}

The accuracy of the LHC-TeV MWWG simulation in determining $m_W$ shifts is studied using PDF variations from the 
ATLAS measurement.  With 28 measurement categories and 25 CT10 PDF eigensets, a statistically accurate comparison 
is made between the emulated measurement procedure and the results of the ATLAS measurement.  
A root-mean-square spread of 1.5 MeV is found between the published and emulated shifts in the various categories and eigensets.
The differences dominantly reflect approximations in the {\textsc{Powheg}}-based reweighting procedure compared to the
kinematic reweighting to NNLO-accurate distributions implemented in Ref.~\cite{Aaboud:2017svj}.
Systematic uncertainties are assessed by varying the response and resolution by $\pm 5\%$, and are 1.1 MeV for the
\pTl~fit and 1.2 MeV for the \mT~fit.

%% file: production.tex
The process of $W$-boson production and decay is similar in $pp$ and $p\bar{p}$
collisions, with differences arising mainly in the parton distribution functions.
Different PDF sets use different input data sets and procedures, and the correlation 
between sets cannot be readily calculated.  Thus the combination is performed by 
adjusting the $m_W$ measurements to a common PDF set through the addition of a 
$\delta m_W^{\scriptsize\textrm{PDF}}$ specific to each experimental result.
Events generated with the {\textsc{Wj-MiNNLO}} Monte Carlo are used to
evaluate the corresponding PDF uncertainty and correlations.  A separate shift
$\delta m_W^{\scriptsize\textrm{pol}}$ is calculated to update the {\textsc{Resbos-C}}
and {\textsc{Resbos-CP}} treatment of the $W$ boson polarization to {\textsc{Resbos2}}.
The line shape of the dilepton invariant mass is also studied, and adjustments are made
for differences in the spectrum due to the CDF generator-level requirements
($\delta m_W^{\scriptsize\textrm{gen}}$) or to the assumed decay width in the measurements 
($\delta m_W^{\mathrm\Gamma}$).  Finally, correlations are estimated for uncertainties
due to electroweak corrections such as final-state photon radiation.

%% file: wpt.tex
In the region relevant to the $m_W$ measurement, the \pTW~distribution is described by a combination of 
perturbative fixed-order QCD, soft-gluon resummation, and non-perturbative effects.  The Tevatron experiments 
use analytical resummation as implemented in {\textsc{ResBos-C} and \textsc{ResBos-CP}, while ATLAS and LHCb
use the \textsc{Pythia8} parton shower interfaced to \textsc{Powheg}.  

Non-perturbative effects influence the very low boson \pTW~region, typically $\pTW<5$~GeV, and are generally 
assumed to be universal between $W$ and $Z$ production (up to differences in $\sqrt{s}$).  In the absence of 
precise direct measurements of the $W$-boson \pT~distribution, all measurements use $Z$-boson data to constrain
the non-perturbative parameters.  The resulting model is then used for the $W$-boson \pT~distribution.  The
associated uncertainty accounts for the limited precision of the $Z$-boson data and for differences between
the $Z$- and $W$-boson production mechanisms, in particular related to the different initial-state parton 
configurations.  

To describe the \pTW~distribution, ATLAS and LHCb tune the shower and non-perturbative parameters in \textsc{Pythia} 
(intrinsic $k_{\scriptsize\textrm{T}}$ and $\alpha_{\scriptsize\textrm{S}}$), and LHCb adds an
$\alpha_{\scriptsize\textrm{S}}$ tune in
\textsc{Powheg}.  The ATLAS tunes use the \pTZ~distribution while the LHCb tunes use an angular distribution in
$Z\to\mu\mu$ decays as well as the $q / \pT$ distribution used for the $m_W$ fit.  CDF fits the non-perturbative
resummation parameters $g_1,g_2$ in \textsc{ResBos-C}
using the \pTZ~distribution, and D0 uses the default values of these parameters in \textsc{ResBos-CP}.  CDF additionally
constrains the region above the peak with a fit for $\alpha_{\scriptsize\textrm{S}}$.  The resulting Tevatron and ATLAS
\pTW~distributions, after event selection and using the detector simulations described in Section~\ref{sec:simulation},
are shown in Figure~\ref{fig:ptwmodels}.

Theoretical uncertainties in the extrapolation from the \pTZ~distribution to the \pTW~distribution
are considered by the ATLAS and CDF experiments, which use the observed $W$-boson \pT~distribution to validate 
(ATLAS) or further constrain (CDF) the associated uncertainty in situ.  
CDF (D0) quotes an uncertainty due to the $W$-boson \pT~modelling of 2.2 (2.4)~MeV and ATLAS quotes 6.0~MeV.
For LHCb an 11 MeV uncertainty is assessed using the envelope of fit results from \textsc{Pythia8} 
(without \textsc{Powheg}), \textsc{Powheg} matched to \textsc{Pythia8} or \textsc{Herwig}, and \textsc{Herwig}
with its own matrix-element calculation.  Since the $W$-boson \pT~distributions are modelled with different generators
or parameter values between the experiments, the corresponding uncertainties are taken to be uncorrelated.  

\begin{figure}
  \centering
  \includegraphics[width=0.9\columnwidth]{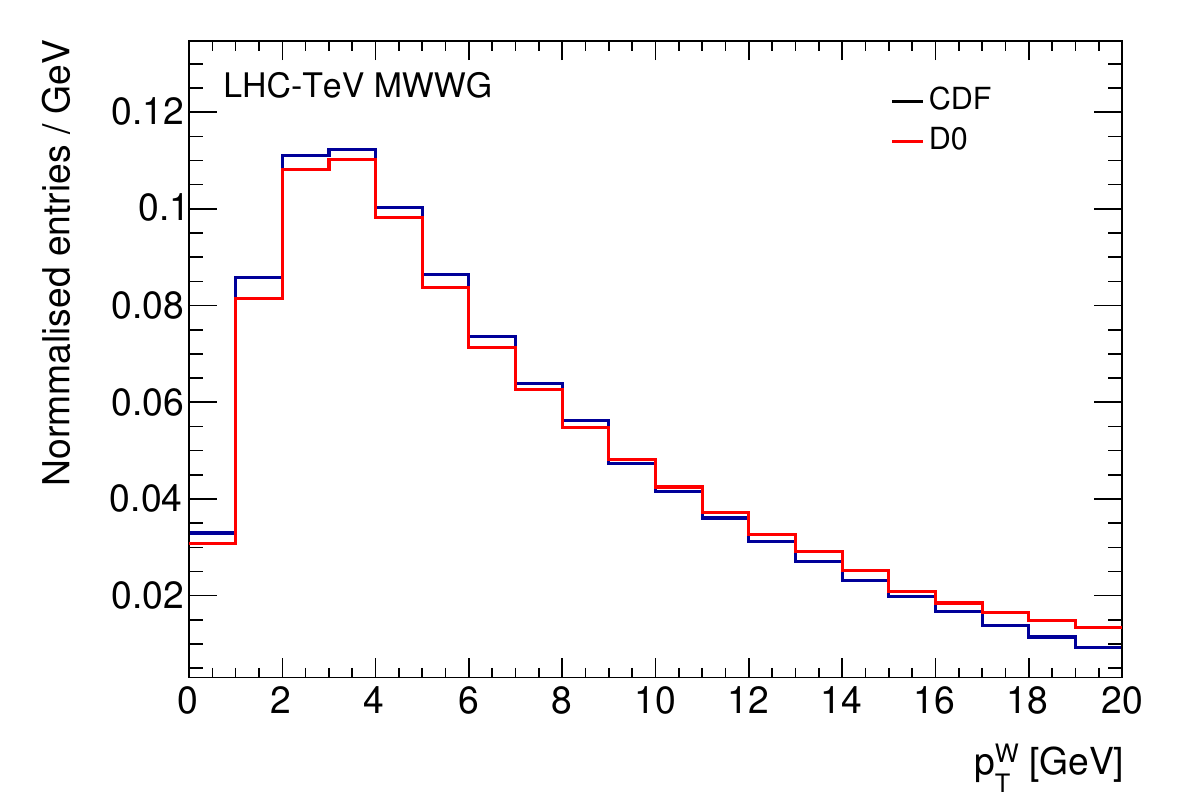}
  \includegraphics[width=0.9\columnwidth]{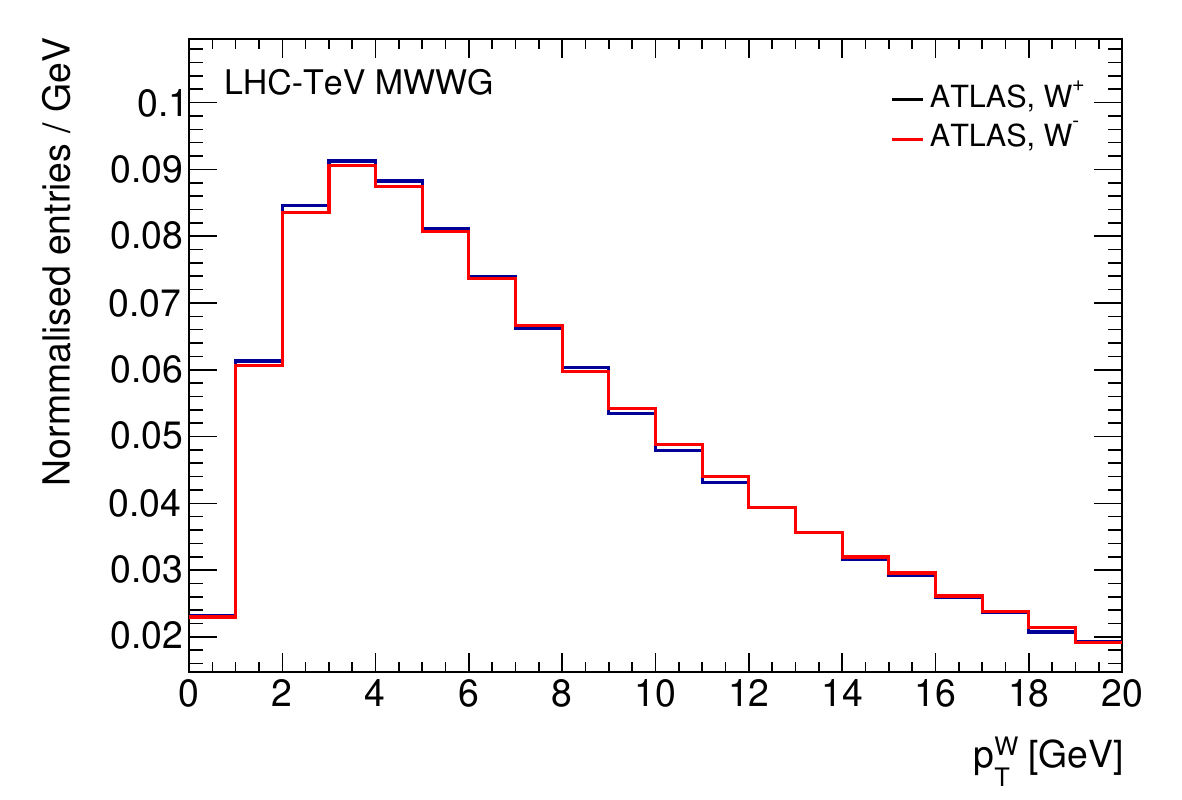}
  \caption{Distributions of generated \pTW for $W^\pm$ candidate events in $p\bar{p}$ collisions at CDF and D0 (top), 
and for $W^+$ and $W^-$ events at ATLAS (bottom).  The distributions represent the best-fit model resulting from the 
analysis of $W$- and $Z$-boson data in the respective experiments, and are shown after all event selection requirements. 
\label{fig:ptwmodels}}
\end{figure}

%% file: pdfextrap.tex
The $\delta m_W^{\scriptsize\textrm{PDF}}$ shift for each PDF set is evaluated for the
Tevatron experiments and ATLAS using the precise NNLO {\textsc{Wj-MiNNLO}} calculation.  
The resulting shifts are are compared to those from the NLO
{\textsc{W\_ew-BMNNP}} and {\textsc{Resbos}} calculations, and the differences are typically
within the statistical precision of the comparison.  For LHCb the PDF shift is determined
with a direct fit to the data as in the original measurement.

All experimental measurements include an {\it in situ} constraint on \pTZ~and/or \pTW.  We
preserve these constraints by reweighting the relevant boson \pT distribution for each PDF
set to match that used in the measurement.  For the Tevatron experiments \pTW~is reweighted,
while in the case of ATLAS \pTZ~is reweighted since the lower recoil resolution does not
provide a significant \pTW~constraint from the data.  For LHCb a constraint on \pTW~is applied
as part of the direct fit to the data for each PDF set.

For each PDF set $\delta m_W^{\scriptsize\textrm{PDF}}$ is evaluated using a common boson $\pT$~distribution
across PDFs separately for each experiment.  For the Tevatron experiments the $W$ boson $\pT$~is
reweighted to match that of the original measurement, due to the observed agreement between the measured
recoil distribution and the model.  In the case of ATLAS the $Z$-boson $\pT$~is reweighted to match the
original measurement, since the lower recoil resolution does not provide a significant $W$-boson
$\pT$ constraint from the data.
For LHCb the PDF shift is determined with a direct fit to the
data as in the original measurement, including constraints on $\pTW$ and the most relevant polarization
coefficient.

In order to facilitate the evaluation of uncertainty correlations, Hessian eigenvector sets are 
used.  The upper and lower uncertainties are taken to be 
\begin{eqnarray}
\sigma_{m_{W}^+} = \left[\sum_i \left({\sigma_{m_{W}^i}}\right)^2\right]^{1/2} \textrm{ if } \sigma_{m_{W}^i}>0 
~\,\,\,\textrm{and} \,\,\,\,\,\, \nonumber \\
\sigma_{m_{W}^-} = \left[\sum_i \left({\sigma_{m_{W}^i}}\right)^2\right]^{1/2} \textrm{ if } \sigma_{m_{W}^i}<0,
\end{eqnarray}
where $i$ runs over the uncertainty sets.  The uncertainties are symmetrized according to 
$\sigma_{m_{W}} = (\sigma_{m_{W}^+} + \sigma_{m_{W}^-})/2$.  For CTEQ PDF sets the translation from 90\% C.L. 
to 68\% C.L. assumes a gaussian distribution, i.e. a division by 1.645.  The effect of each PDF eigenset 
is correlated across experiment or measurement category, and its contribution to the covariance between 
any two measurements $\alpha$ and $\beta$ is given by
$C^i_{\alpha\beta} = \sigma_{m_{W\alpha}^i} \sigma_{m_{W\beta}^i}$.
Accounting for all eigensets of a given PDF, the total uncertainty covariance and the corresponding 
uncertainty correlation are calculated as
\begin{equation}
C^{\scriptsize\textrm{PDF}}_{\alpha\beta} = \sum_i C^i_{\alpha\beta},~\,\,\,\textrm{and}\,\,\,\,\rho_{\alpha\beta} = 
\frac{\sum_i \sigma_{m_{W\alpha}^i} \sigma_{m_{W\beta}^i}}{\sigma_{m_{W\alpha}}\sigma_{m_{W\beta}}}.
\end{equation}

Tables~\ref{tab:shiftmT} and \ref{tab:shiftpTlmet} show $\delta m_W^{\scriptsize\textrm{PDF}}$ for each PDF set and each 
experiment using distributions based on the transverse mass and the lepton or neutrino $p_T$, respectively.  
For simplicity the ATLAS shifts are shown inclusively in lepton $\eta$, though separated by boson charge.
The PDF uncertainties for each measurement are shown in Table~\ref{tab:pdfunc}, and the correlation 
matrices for the the most recent PDF sets are shown in Fig.~\ref{fig:pdfcorr}.  Correlation matrices for
older sets are provided in the Appendix.

\begin{table}[!tbp]
  \centering 
\begin{tabular}{lrrrr}
\hline \hline 
              PDF set  &        D0   &            CDF &  ATLAS $W^+$ &  ATLAS $W^-$ \\
\hline
              CTEQ6    &     $-14.6$ &            0.0 &           -- &           -- \\ 
              CTEQ6.6  &         0.0 &           14.2 &           -- &           -- \\ 
              CT10     &      $-0.5$ &           14.3 &          0.0 &          0.0 \\ 
              CT14     &      $-8.7$ &            5.2 &       $-0.5$ &       $-7.6$ \\ 
              CT18     &      $-7.5$ &            6.5 &         13.4 &       $-5.5$ \\ 
              ABMP16   &     $-17.9$ &         $-2.4$ &      $-25.7$ &       $-7.9$ \\ 
              MMHT2014 &     $-10.1$ &            4.5 &       $-3.6$ &          9.1 \\ 
              MSHT20   &     $-12.9$ &            2.5 &      $-22.3$ &          4.2 \\ 
              NNPDF3.1 &      $-1.0$ &           13.1 &      $-14.6$ &       $-6.3$ \\ 
              NNPDF4.0 &         6.2 &           20.1 &      $-23.3$ &          4.3 \\ 
\hline \hline
\end{tabular}
  \caption{Values of $\delta m_W^{\scriptsize\textrm{PDF}}$ in MeV for each PDF set using the \mT~fit
  distribution, determined using the {\textsc{Wj-MiNNLO}} calculation.
}
\label{tab:shiftmT}
\end{table}

\begin{table*}[!tbp]
  \centering 
\begin{tabular*}{\linewidth}{l @{\extracolsep{\fill}}rrrrrrr}
\hline \hline 
              PDF set  & D0 \pTl & D0 \met & CDF \pTl &    CDF \met     & ATLAS $W^+$ &  ATLAS $W^-$ & LHCb    \\
\hline
              CTEQ6    & $-17.0$ & $-17.7$ &      0.0 &             0.0 &          -- &          -- & -- \\ 
              CTEQ6.6  &     0.0 &     0.0 &     15.0 &            17.0 &          -- &          -- & -- \\ 
              CT10     &     0.4 &  $-1.3$ &     16.0 &            16.3 &         0.0 &         0.0 & -- \\ 
              CT14     &  $-9.7$ & $-10.6$ &      5.8 &             6.8 &        -1.2 &      $-5.8$ & 1.1 \\ 
              CT18     &  $-8.2$ &  $-9.3$ &      7.2 &             7.7 &        12.1 &      $-2.3$ & $-6.0$ \\ 
              ABMP16   & $-19.6$ & $-21.5$ &   $-1.4$ &          $-2.4$ &     $-22.5$ &      $-3.1$ & 7.7 \\ 
              MMHT2014 & $-10.4$ & $-12.7$ &      6.1 &             5.5 &      $-2.6$ &         9.9 & $-10.8$ \\ 
              MSHT20   & $-13.7$ & $-15.4$ &      3.6 &             4.1 &     $-20.9$ &         4.5 & $-2.0$ \\ 
              NNPDF3.1 &  $-1.0$ &  $-1.2$ &     14.0 &            15.1 &     $-14.1$ &        -1.8 & 6.0 \\ 
              NNPDF4.0 &     6.7 &     8.1 &     20.8 &            24.1 &     $-22.4$ &         6.9 & 8.3 \\ 
\hline \hline
\end{tabular*}
  \caption{Values of $\delta m_W^{\scriptsize\textrm{PDF}}$ in MeV for each PDF set using the \pTl~(all experiments)
  or \met~(CDF and D0) distribution,
   determined using the {\textsc{Wj-MiNNLO}} calculation.
}
\label{tab:shiftpTlmet}
\end{table*}

\begin{table}[!tbp]
  \centering 
\begin{tabular}{lrrrr}
\hline \hline
              PDF set  &    D0   &       CDF &           ATLAS &  LHCb    \\
\hline 
              CTEQ6    &     --  &      14.1 &             --  &   --     \\
              CTEQ6.6  &    15.1 &       --  &             --  &   --     \\
              CT10     &     --  &       --  &             9.2 &   --     \\
              CT14     &    13.8 &      12.4 &            11.4 &   10.8   \\
              CT18     &    14.9 &      13.4 &            10.0 &   12.2   \\
              ABMP16   &     4.5 &       3.9 &             4.0 &    3.0   \\
              MMHT2014 &     8.8 &       7.7 &             8.8 &    8.0   \\
              MSHT20   &     9.4 &       8.5 &             7.8 &    6.8   \\
              NNPDF3.1 &     7.7 &       6.6 &             7.4 &    7.0   \\
              NNPDF4.0 &     8.6 &       7.7 &             5.3 &    4.1   \\
\hline \hline 
\end{tabular}
  \caption{Uncertainty in MeV for each PDF set after combining the individual fit categories.  
 }
\label{tab:pdfunc}
\end{table}

 \begin{figure}[!tbp]
   \includegraphics[width=0.9\columnwidth]{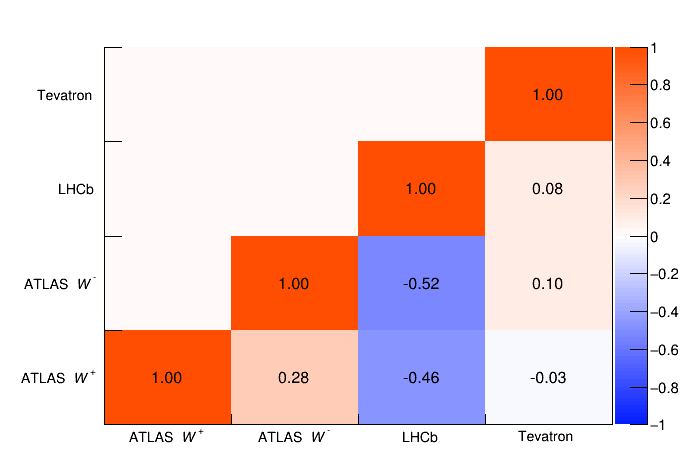}
   \includegraphics[width=0.9\columnwidth]{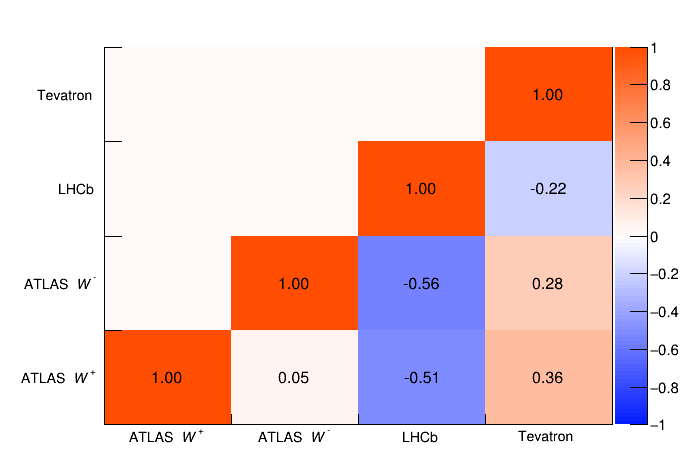}
   \includegraphics[width=0.9\columnwidth]{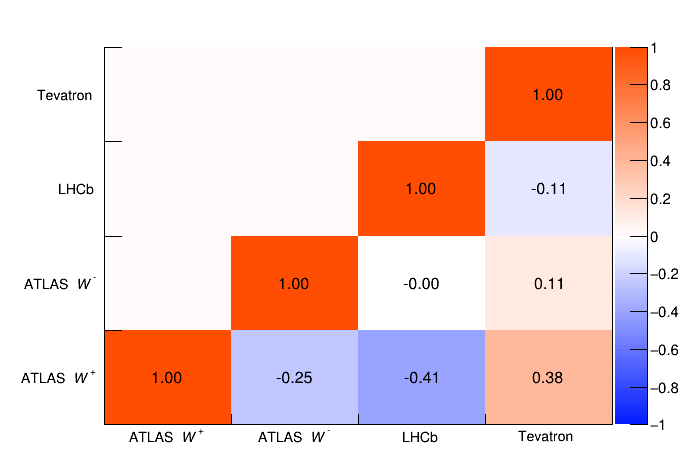}
   \includegraphics[width=0.9\columnwidth]{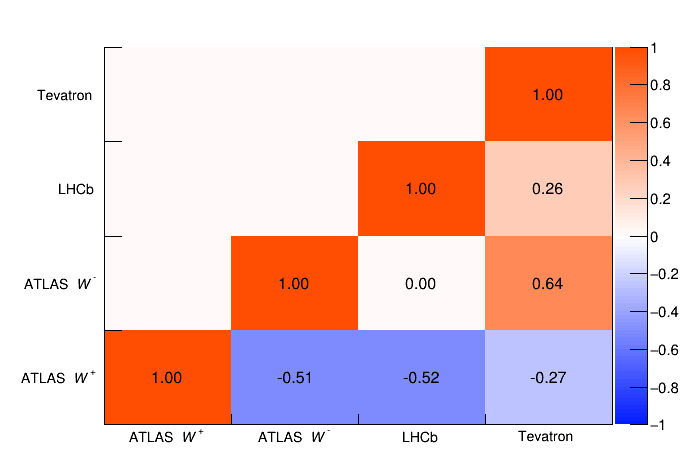}
   \caption{PDF uncertainty correlation matrices for the CT18, MSHT20, NNPDF4.0, and ABMP16 PDF sets, in order from top to bottom. 
}
\label{fig:pdfcorr}
\end{figure}

%% file: pdf.tex
The various kinematic distributions and fiducial regions used to fit $m_W$ in the ATLAS and
Tevatron experiments provides some sensitivity to PDF predictions.  Other $W$- and $Z$-boson 
measurements from the LHC and the 
Tevatron provide more significant PDF constraints and are used in the determination of the 
PDF sets.  This section compares the compatibility of these other measurements with the
various PDF sets.  Some sets have low compatibility and are not favoured for an $m_W$
combination.

The $W$-boson rapidity ($y_W$) distribution affects the $m_W$ measurement through the
\pTl~distribution: more central $W$ bosons can have more forward-decaying leptons within the 
detector acceptance, lowering the mean observed \pTl.  Measurements that probe PDF parameters 
describing $y_W$ include the $Z$ boson rapidity $y_Z$ and the asymmetries in the rapidity 
distribution between positive and negative $W$ bosons ($A_W$), or similarly the positive and 
negative charged leptons from their decay ($A_\ell$).  These measurements are considered in this 
compatibility study, and are shown in Table~\ref{tab:datasets}.  

The comparison between data and predictions is performed with the xFitter~\cite{Bertone:2017tig} 
framework.  A $\chi^2$ measure is constructed including all experimental uncertainties and their 
correlations, as well as the PDF uncertainties.
Theory predictions
are calculated at NNLO in QCD and corrected to NLO electroweak predictions using multiplicative $k$-factors
in each measurement bin.  PDF uncertainties are computed at NLO in QCD using Applgrids~\cite{Carli:2010rw}
with calculations from MCFM-6.8~\cite{Campbell:1999ah}.
The results for various PDF sets are shown in Table~\ref{tab:pdfchi2}.

Most of the Drell-Yan measurements have good $\chi^2$ values for all PDFs.  The most significant 
outlier is the D0 $W\rightarrow e \nu$ lepton asymmetry measurement, for which the CT18 set has 
the lowest $\chi^2$ primarily due to its larger uncertainties.  These larger uncertainties also
reduce the correlated $\chi^2$, which represents the contribution from correlated
uncertainties~\cite{alekhin:2015herafitter}.
The correlated $\chi^2$ reduces from 251 to 43 after including
PDF uncertainties in the CT18 set; the corresponding reduction for the NNPDF3.1 set is 110 to 76.
The overall probability of consistency of the combined datasets is 1.5\% for the CT18 set, and is
much lower for the other sets.
Among the studied PDF sets CT18 is therefore considered to give the most accurate estimate of the
68\% C.L. interval for combined $W$- and $Z$-boson measurements.

\begin{table}[!tbp]
  \begin{center}
    \begin{tabular}{cccllc}
\hline \hline
Exp.                       & Obs.     & Decay    & $\sqrt{s}$ & Lum.           & bins \\
\hline
CDF~\cite{CDF:2009cjw}     & $A_W$    & $e\nu$   & 1.96~TeV   & 1~fb$^{-1}$    & 13 \\ 
CDF~\cite{CDF:2010vek}     & $y_Z$    & $ee$     & 1.96~TeV   & 2.1~fb$^{-1}$  & 28 \\ 
\hline
D0~\cite{D0:2007djv}       & $y_Z$    & $ee$     & 1.96~TeV   & 0.4~fb$^{-1}$  & 28 \\ 
D0~\cite{D0:2013xqc}       & $A_\ell$ & $\mu\nu$ & 1.96~TeV   & 7.3~fb$^{-1}$  & 12 \\ 
D0~\cite{D0:2014kma}       & $A_\ell$ & $e\nu$   & 1.96~TeV   & 9.7~fb$^{-1}$  & 13 \\ 
\hline
ATLAS~\cite{ATLAS:2016nqi} & $Z$,$W$ & $\ell\ell$,$\ell\nu$ & 7~TeV & 4.7~fb$^{-1}$ & 61 \\ 
\hline \hline
\end{tabular}
    \caption{Drell-Yan measurements used for the PDF compatibility study.
    } 
      \label{tab:datasets} 
\end{center}
\end{table}

\begin{table*}[!tbp]
  \begin{center}
  \begin{tabular*}{\linewidth}{l @{\extracolsep{\fill}}lllllll}
\hline \hline 
  Measurement                & NNPDF3.1   & NNPDF4.0 & MMHT14   & MSHT20  & CT14      & CT18      & ABMP16  \\
\hline 
  CDF $y_Z$                  & 24 / 28   & 28 / 28   & 30 / 28  & 32 / 28 & 29 / 28   & 27 / 28   & 31 / 28  \\
  CDF $A_W$                  & 11 / 13   & 14 / 13   & 12 / 13  & 28 / 13 & 12 / 13   & 11 / 13   & 21 / 13  \\
  D0  $y_Z$                  & 22 / 28   & 23 / 28   & 23 / 28  & 24 / 28 & 22 / 28   & 22 / 28   & 22 / 28  \\
  D0 $W \to e\nu$ $A_\ell$   & 22 / 13   & 23 / 13   & 52 / 13  & 42 / 13 & 21 / 13   & 19 / 13   & 26 / 13  \\
  D0 $W \to \mu\nu$ $A_\ell$ & 12 / 10   & 12 / 10   & 11 / 10  & 11 / 10 & 11 / 10   & 12 / 10   & 11 / 10  \\
  ATLAS peak CC $y_Z$        & 13 / 12   & 13 / 12   & 58 / 12  & 17 / 12 & 12 / 12   & 11 / 12   & 18 / 12  \\
  ATLAS $W^{-}$ $y_\ell$        & 12 / 11   & 12 / 11   & 33 / 11  & 16 / 11 & 13 / 11   & 10 / 11   & 14 / 11  \\
  ATLAS $W^{+}$ $y_\ell$        &  9 / 11   &  9 / 11   & 15 / 11  & 12 / 11 & 9 / 11    &  9 / 11   & 10 / 11  \\
  Correlated $\chi^2$        & 75        & 62        & 210      & 88      & 81        & 41        & 83       \\
\hline
  Total $\chi^2$ / d.o.f.    & 200 / 126 & 196 / 126 & 444 / 126 & 270 / 126 & 210 / 126 & 162 / 126  & 236 / 126  \\
\hline
  p($\chi^2,n$)        & 0.003\%   & 0.007\%   & $<10^{-10}$  & $<10^{-10}$ & $0.0004\%$ & 1.5\% & $10^{-8}$   \\
\hline \hline 
    \end{tabular*}
    \caption{$\chi^2$ per degree of freedom for the Tevatron $Z$-rapidity and $W$- and $l$-asymmetry
    measurements at $\sqrt{s}=1.96$~TeV, and the LHC $Z$-rapidity and $W$ lepton-rapidity measurements at $\sqrt{s}=7$~TeV.  
    The total $\chi^2$ is the sum of those quoted for individual measurements along with a separate contribution for
    correlated uncertainties, where the latter is extracted using a nuisance parameter representation of the
    $\chi^2$~\cite{alekhin:2015herafitter}.  The CT14 and CT18 PDF uncertainties correspond to 68\% coverage, obtained by
    rescaling the eigenvectors by a factor of 1/1.645.  The probability of obtaining a total $\chi^2$ at least as high
    as that observed is labelled p($\chi^2,n$).
}
    \label{tab:pdfchi2}
  \end{center}
\end{table*}

%% file: polarization.tex
The $W$-boson polarization affects the lepton decay angles, and in turn the transverse momentum of the leptons.
A general expression for the fully differential distribution of the charged lepton is 
\begin{eqnarray}
\frac{d\sigma}{d\pTW dy dm d\Omega}  & = & \frac{d\sigma}{d\pTW dy dm} [(1 + \cos^2\theta) \nonumber \\
 & + &  \frac{1}{2} A_0 (1-3\cos^2\theta) + A_1 \sin 2\theta \cos\phi  \nonumber \\
 & + & \frac{1}{2}A_2 \sin^2\theta\cos 2\phi + A_3 \sin\theta\cos\phi \nonumber \\
 & + & A_4 \cos\theta + A_5 \sin^2\theta\sin 2\phi \nonumber \\
 & + & A_6 \sin 2\theta\sin\phi + A_7 \sin\theta\sin\phi ],
\label{eq:wai}
\end{eqnarray}
where the decay angles $\theta, \phi$ are expressed in the Collins-Soper (C-S) frame~\cite{Collins:1977iv},
and the $A_i$ coefficients depend on the \pT, rapidity, and invariant mass of the $\ell\nu$ system.  The 
coefficients can be calculated perturbatively in $\alpha_{\mathrm{S}}$, with $A_5,~A_6$, and $A_7$ becoming non-zero
only at NNLO in QCD.  The $A_0$ term primarily reflects the relative fractions of the $qq\to W$, $qg\to Wq$, and
higher-order subprocesses, and has a significant \pTW~dependence while being nearly independent of boson rapidity.
The $A_4$ term produces a forward-backward asymmetry, and is thus sensitive to the directions of the incoming
quark and anti-quark in the dominant $q\bar{q}' \to W$ process. It depends on rapidity and on the PDF set used
in the calculation, and decreases with increasing \pTW. 

\begin{figure}[!tbp]
  \centering
  \includegraphics[width=0.88\columnwidth]{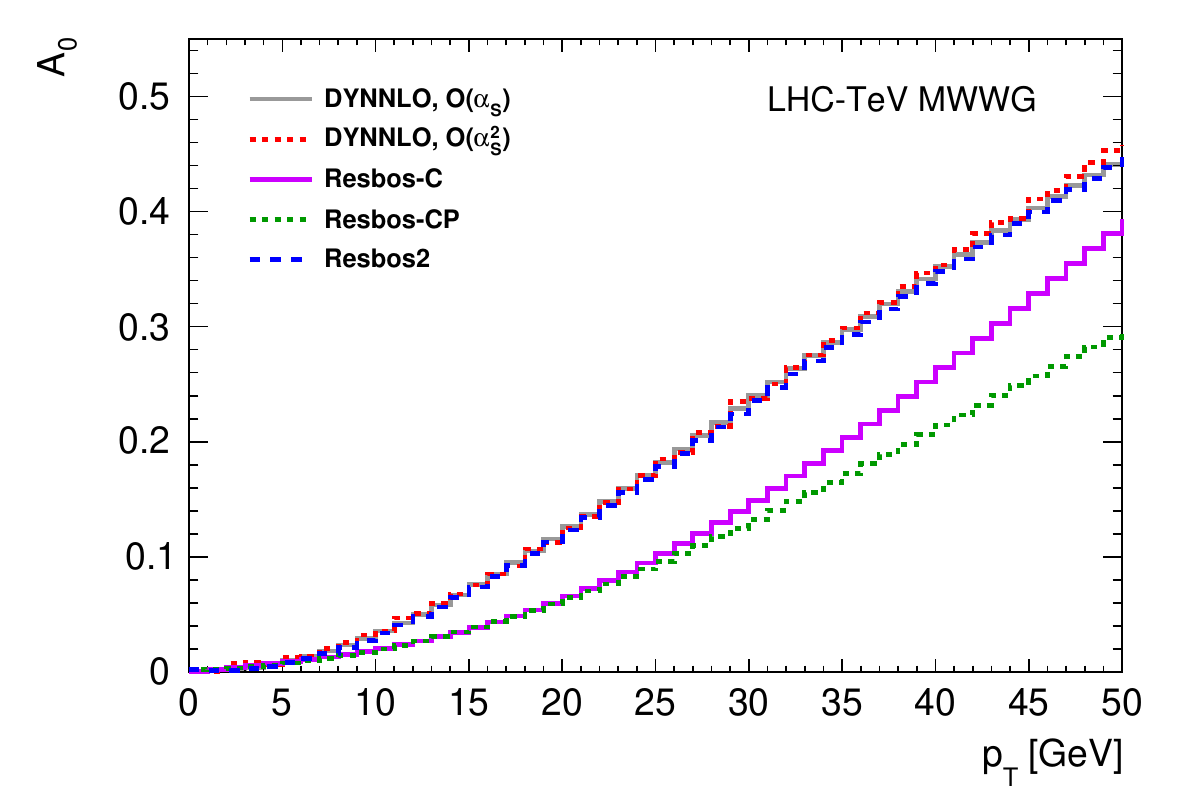}
  \includegraphics[width=0.88\columnwidth]{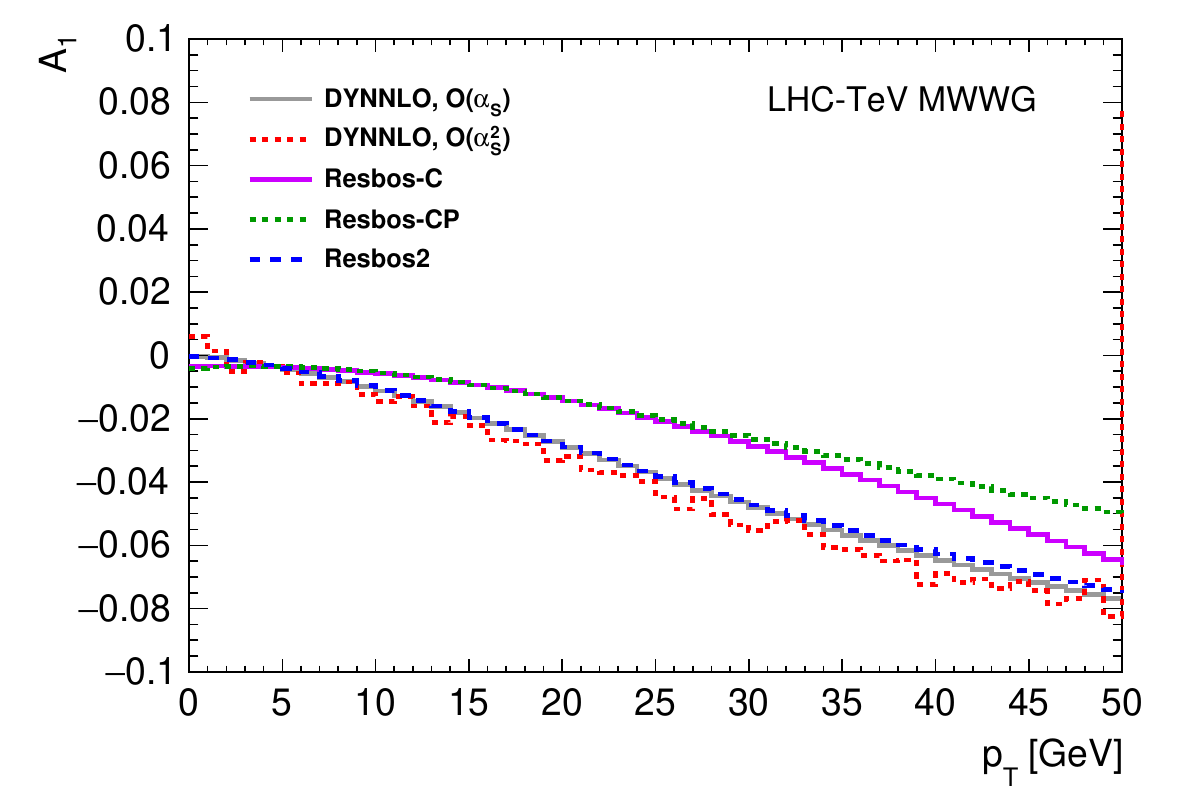}
  \includegraphics[width=0.88\columnwidth]{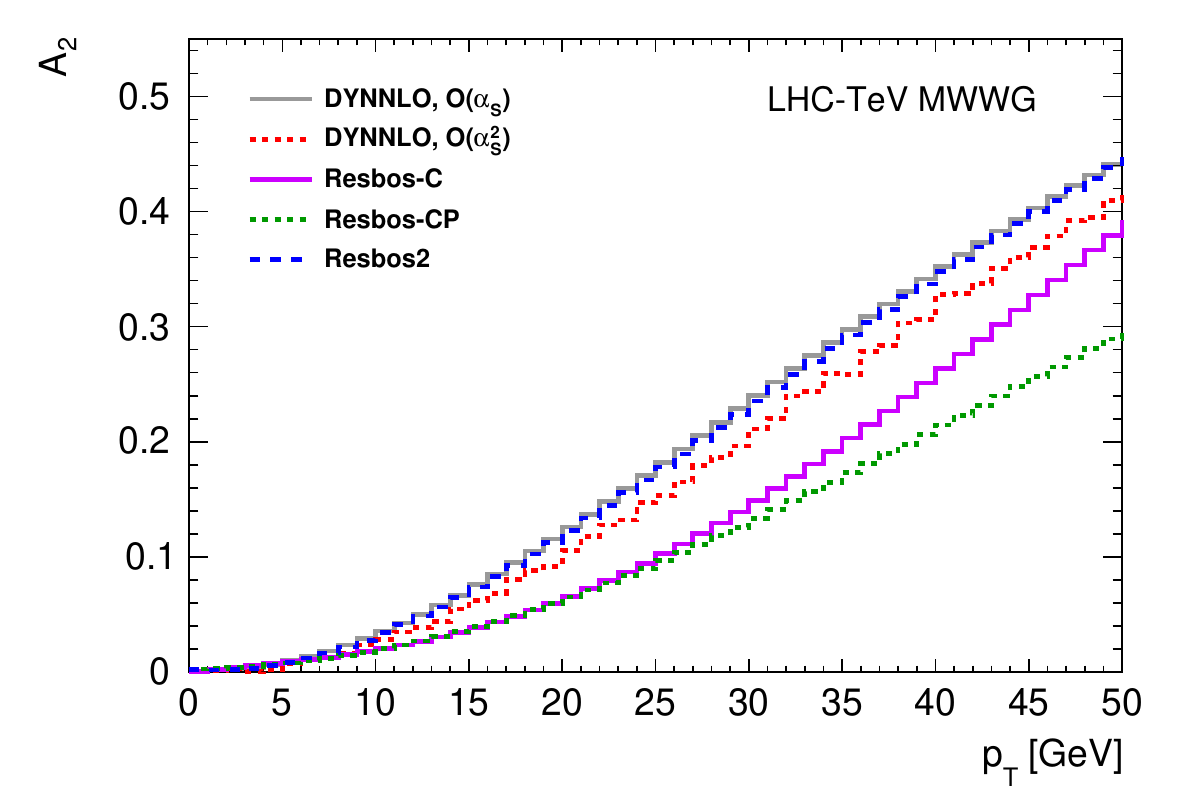}
  \includegraphics[width=0.88\columnwidth]{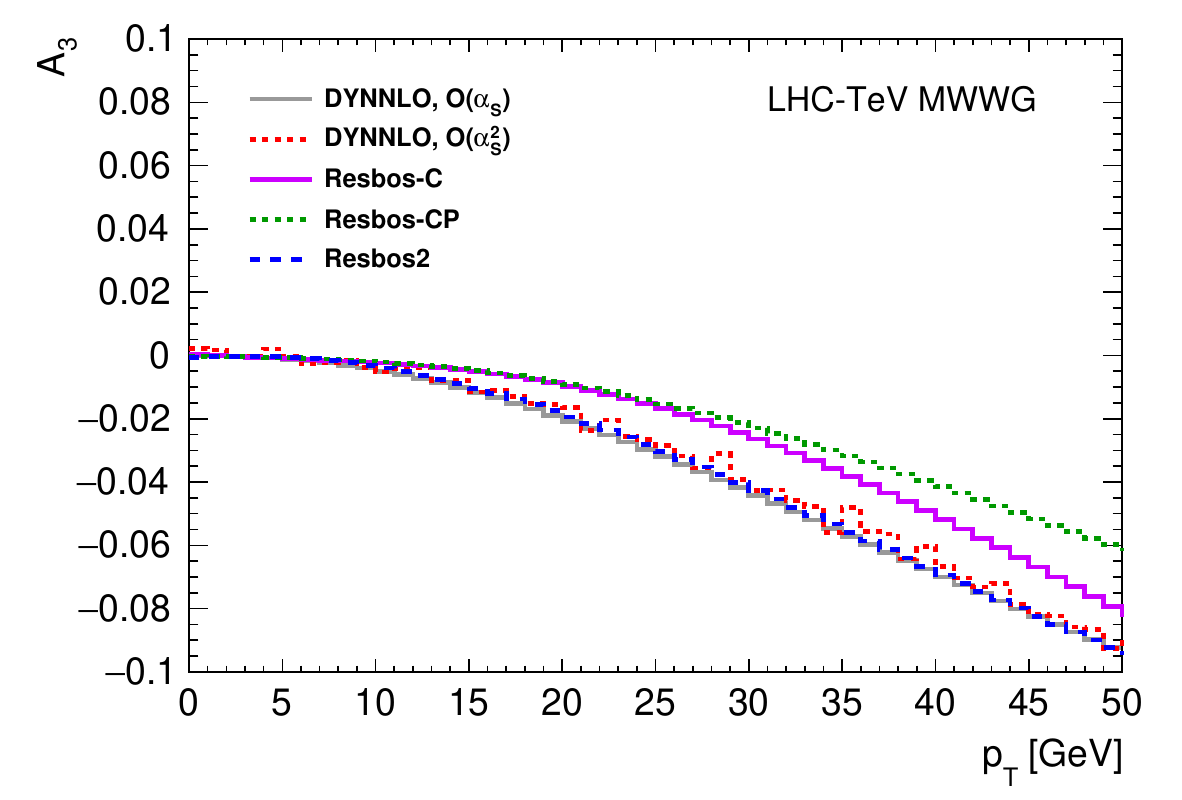}
  \caption{$A_0$ to $A_3$ as a function of \pTW~extracted from 
    \textsc{ResBos-C}, \textsc{ResBos-CP}, \textsc{ResBos2}, \textsc{DYNNLO} at
    ${\cal O}(\alpha_{\scriptsize\textrm{S}})$, and \textsc{DYNNLO} at ${\cal O}(\alpha^2_{\scriptsize\textrm{S}})$
    in $p\bar{p}$ collisions at 1.96 TeV.  The CTEQ6M PDF set is used for all generators except \textsc{ResBos-CP},
    for which CTEQ6.6 is used.  The \textsc{ResBos-C} and \textsc{ResBos2} calculations are at
    ${\cal O}(\alpha_{\mathrm{S}})$ in QCD, and \textsc{ResBos-CP} is at ${\cal O}(\alpha^2_{\scriptsize\textrm{S}})$.
      The difference between DYNNLO at ${\cal O}(\alpha_{\scriptsize\textrm{S}})$ and
      ${\cal O}(\alpha^2_{\scriptsize\textrm{S}})$ has an ${\cal{O}}$(1 MeV) effect on $\delta m_W^{\mathrm{pol}}$.
  }
\label{fig:aiplots}
\end{figure}

The \textsc{ResBos-C} and \textsc{ResBos-CP} codes resum a subset of contributions to Equation~\ref{eq:wai},
specifically those affecting the $(1 + \cos^2\theta)$ and $A_4 \cos\theta$ terms.  This partial resummation modifies
the $A_0$--$A_3$ terms relative to fixed-order predictions, as demonstrated in Figure~\ref{fig:aiplots}, where
$A_0-A_3$ are shown for $W$-boson events generated at $\sqrt{s}=1.96$~TeV with \textsc{ResBos-C}, \textsc{ResBos-CP},
\textsc{ResBos2}, and \textsc{DYNNLO}.  The partial-resummation predictions differ with respect to measurements
performed at the LHC~\cite{STDM-2014-10}, which instead agree with fully-resummed calculations such as
\textsc{ResBos2} or {\sc Wj-MiNNLO}, and fixed-order calculations such as \textsc{DYNNLO}.  

Experimental fits for $m_W$ in data use theoretical predictions of the leptonic angular distributions from
\textsc{ResBos-C} for CDF, \textsc{ResBos-CP} for D0, \textsc{DYNNLO}~\cite{Catani:2007vq,Catani:2009sm} for
ATLAS, and DYTurbo for LHCb.  The CDF experiment applies a post-fit correction to reproduce the NNPDF3.1 PDF
prediction, and this correction includes the effect of updating the angular coefficients to those calculated
by \textsc{ResBos2}.  

\begin{figure}[!htbp]
  \centering
  \includegraphics[width=0.9\columnwidth]{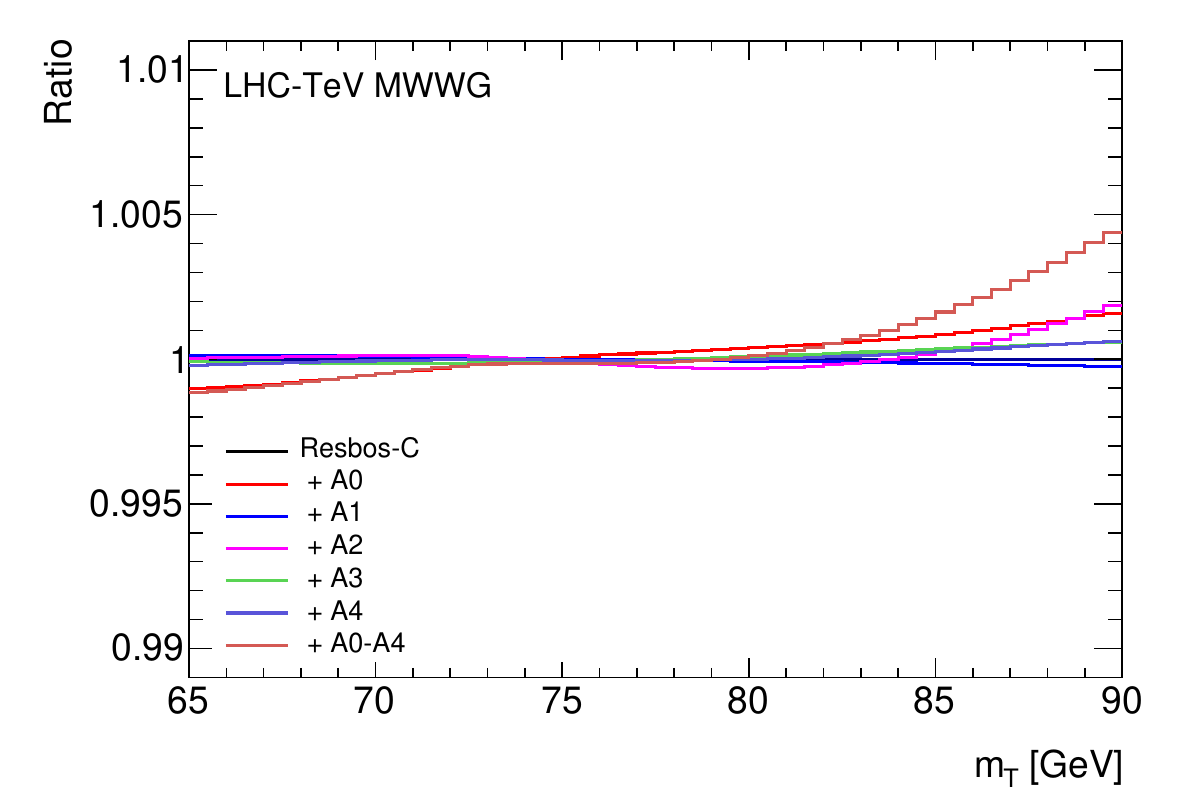}
  \includegraphics[width=0.9\columnwidth]{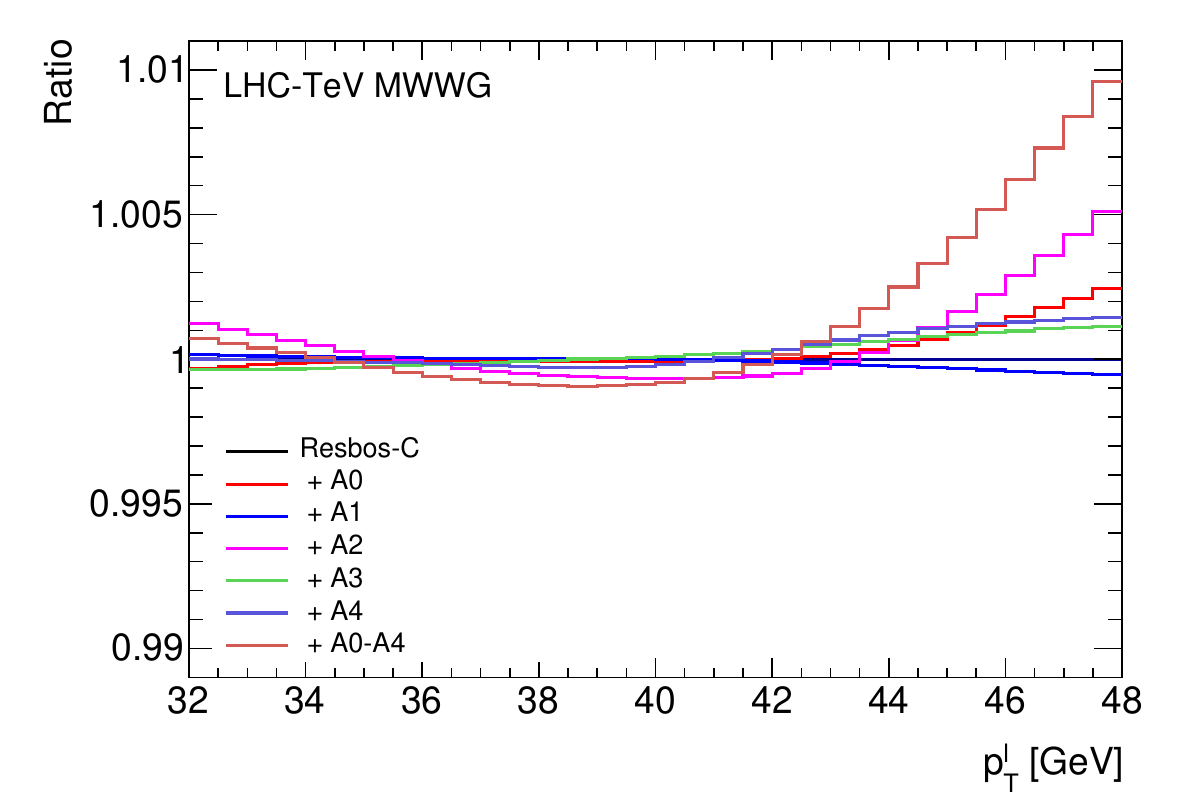}
  \caption{Relative effect of reweighting the $A_0$--$A_4$ coefficients from \textsc{ResBos-C} to 
\textsc{ResBos2} on the CDF \mT~(left) and \pTl~(right) distributions. 
  }
\label{fig:airewCDF}
\end{figure}

In order to achieve a common theoretical treatment of the $W$-boson polarization, the results of the CDF and 
D0 fits to the measurement distributions are adjusted to correspond to the \textsc{ResBos2} calculation of the 
leptonic angular distributions at ${\cal O}(\alpha_{\mathrm{S}})$.  Events generated with \textsc{ResBos-C} or
\textsc{ResBos-CP} 
are reweighted such that the $A_0-A_4$ coefficients match those of \textsc{ResBos2}, as functions of \pTW~and 
$y_W$.  The $W$-boson \pT~is fixed to that of the original measurement, in the same manner as for the 
$\delta m_W^{\scriptsize\textrm{PDF}}$ evaluations in Sec.~\ref{sec:corrPDF}.  The impact of the reweighting on
the CDF \mT~and \pTl~distributions is shown in Fig.~\ref{fig:airewCDF}, and the $\delta m_W^{\mathrm{pol}}$
values from reweighting
the $A_i$ coefficients individually and together are given in Tables~\ref{tab:aireweightcdf}~and~\ref{tab:aireweightd0}
for CDF and D0, respectively.  The reweighting procedure reproduces the direct fit from \textsc{ResBos-C} or
\textsc{ResBos-CP} to \textsc{ResBos2}, as expected since the basis of spherical harmonics is complete and exact.
The results of the reweighting procedure for the D0 
configuration, $\delta m_W^{\mathrm{pol}} = -6.4$, $-6.9$, and $-15.8$~MeV for the \mT, \pTl, and
\met~distributions, respectively, are applied to the measured $m_W$.  For CDF, values of
$\delta m_W^{\mathrm{pol}} = -9.5$, $-8.4$, and $-12.5$~MeV for the \mT, \pTl, and
\met~distributions, respectively, are applied to events generated with \textsc{ResBos-C}.

ATLAS estimates a 5.8 MeV polarization modelling uncertainty based on the precision of measurements on the 
$Z$-boson resonance, while the LHCb uncertainty of 10 MeV arises from its determination of the $A_3$ 
coefficient as part of its fit for $m_W$.  These uncertainties are taken to be uncorrelated.  The Tevatron 
experiments do not include a corresponding uncertainty in their measurements.  An uncorrelated uncertainty 
is applied to the shift calculated for each experiment to account for the limitations of the parameterized MWWG
simulation.  This uncertainty is $\approx 1$~MeV and is similar to that obtained by taking the difference between
the NLO and NNLO fixed-order calculations of the leptonic angular coefficients.

\begin{table}[!tbp]
  \centering
  \begin{tabular}{lccc}
    \hline \hline
    Coefficient      &  \mT             & \pTl           & \pTnu   \\
    \hline
    $A_0$            &  $-6.3$          & $-2.6$         & $-9.1$     \\
    $A_1$            &   1.1            &  1.3           &  0.3      \\
    $A_2$            &  $-0.7$          &  0.4           & $-3.2$      \\
    $A_3$            &  $-2.1$          & $-4.1$         &  1.0       \\
    $A_4$            &  $-1.4$          & $-3.3$         & $-1.6$        \\
    $A_0-A_4$        &  $-9.5$          & $-8.4$         & $-12.5$        \\
    \hline
    \textsc{ResBos2} &  $-10.2 \pm 1.1$ & $-7.6 \pm 1.2$ & $-11.8 \pm 1.4$ \\
    \hline 
    Difference       &  $-0.7 \pm 1.1$  &  $0.8 \pm 1.2$ &   $0.7 \pm 1.4$ \\
    \hline \hline
  \end{tabular}
  \caption{Values of $\delta m_W^{\mathrm{pol}}$ in MeV associated with reweighting each $A_i$ coefficient from
    {\textsc{Resbos-C}} to {\textsc{Resbos2}} for the CDF detector, as well as the result of a direct fit
    to {\textsc{ResBos2}}.  The result of the direct fit is consistent with that of the reweighting. }
\label{tab:aireweightcdf}
\end{table}

\begin{table}[!tbp]
  \centering
  \begin{tabular}{lccc}
    \hline \hline
    Coefficient      &  \mT         & \pTl     & \pTnu \\
    \hline
    $A_0$            &  $-9.8$      & $-7.3$   & $-15.6$ \\
    $A_1$            &   1.9        &  2.4     &   1.8 \\
    $A_2$            &   3.0        &  3.3     &  $-2.7$ \\
    $A_3$            &  $-1.6$      & $-2.9$   &   0.4 \\
    $A_4$            &   0.2        & $-2.3$   &   0.5 \\
    $A_0-A_4$        &  $-6.4$      & $-6.9$   & $-15.8$ \\
    \hline
    \textsc{ResBos2} & $-7.8 \pm 1.0$ & $-6.6 \pm 1.1$ & $-16.5 \pm 1.2$ \\
    \hline 
    Difference       & $-1.4 \pm 1.0$ & $0.3 \pm 1.1$  & $-0.7 \pm 1.2$ \\
    \hline \hline
  \end{tabular}
  \caption{Values of $\delta m_W^{\mathrm{pol}}$ in MeV associated with reweighting each $A_i$ coefficient from
    {\textsc{ResBos-CP}} to {\textsc{Resbos2}} for the D0 detector, as well as the result of a direct fit to 
{\textsc{ResBos2}}.  The result of the direct fit is consistent with that of the reweighting. }
\label{tab:aireweightd0}
\end{table}

%% file: invmass.tex
The details of the event generation for the $m_W$ measurement differ between measurements due to 
the assumed $W$-boson width ${\mathrm \Gamma}_W$ and to a restriction on the generated $\ell\nu$ invariant 
mass range in the CDF sample.  These lead to $\delta m_W$ corrections on the direct fits to these 
samples.  

The assumed ${\mathrm \Gamma}_W$ values used by the experiments are as follows: 2089.5 MeV for the CDF 
measurement; 2099 and 2100 MeV for the D0 measurements with 1.0~fb$^{-1}$ and 4.3~fb$^{-1}$, 
respectively; 2094 MeV for the ATLAS measurement; and 2085 for the LHCb measurement.  Using the  
SM prediction of ${\mathrm \Gamma}_W = 2089.5 \pm 0.6$~MeV leads to $\delta m_W^{\mathrm \Gamma}=0.0$, 1.4,
1.5, 0.7, and $-0.7$~MeV for the CDF, D0 1.0 fb$^{-1}$, D0 4.3 fb$^{-1}$, ATLAS, and LHCb measurements,
respectively.

The {\textsc{ResBos-C}} events used in the fit to the CDF data includes an $\ell\nu$ invariant mass 
requirement of $m_{\ell\nu}<150$~GeV.  Differences of up to 10\% are observed between {\textsc{Resbos-C}} 
and {\textsc{ResBos2}} for an invariant mass below 70~GeV, though these have a negligible effect on the 
measurement.  Using the {\textsc{ResBos2}} invariant mass distribution without any requirements leads 
to $\delta m_W^{\scriptsize\textrm{gen}} = -1.6, -3.4,$ and $-3.2$~MeV for the \mT, \pTl, and \met~distributions,
respectively, for the CDF fit results.  The measured CDF $m_W$ accounts for these effects as part of the update
of the PDF set to NNPDF3.1.  Smaller differences are observed between {\textsc{Resbos-CP}} and {\textsc{ResBos2}}, 
and there is no significant $\delta m_W^{\scriptsize\textrm{gen}}$ from the invariant mass modelling for the D0
measurement.

%% file: ewk.tex
The dominant electroweak effect on the $m_W$ measurement is final-state QED radiation~\cite{EWK},
which reduces the momentum of the charged lepton from the $W$-boson decay.
The experiments model this radiation using generators that resum multiple soft photon
emissions above an energy threshold. Uncertainties on the modelling of electroweak
corrections include:
(1) the perturbative calculation of photon radiation, including the modelling of single-photon and multi-photon
emission and the matching of the fixed-order and all-orders descriptions
(2) the energy threshold for producing final-state photons; and 
(3) higher-order corrections from final-state $e^+ e^-$ pair production.
Tables~\ref{tbl:ewkelectrons} and  \ref{tbl:ewkmuons} list the size of these uncertainties
for each experiment in the electron and muon channels, respectively.
The uncertainties are completely correlated between the decay channels.

\begin{table}[!tp]
  \centering 
  \begin{tabular}{lccc}
  \hline \hline 
     Uncertainty        & CDF       & D0     & ATLAS     \\ 
     \hline 
    Perturbative photon rad.   & 2.3 (2.3) & 5 (5) & 2.5 (3.3) \\ 
    Photon energy cutoff  & 1 (1)  & 2 (1) & ---    \\
    FSR $e^+e^-$       & 1 (1)     & ---  & 0.8 (3.6) \\
    \hline 
    Total              & 2.7 (2.7)     & 7 (7) & 2.6 (4.9) \\
    \hline \hline 
  \end{tabular}
  \caption{QED uncertainties in MeV on the $m_W$ measurement in the electron 
  channel using the \mT~(\pT) fit.  The uncertainties are uncorrelated except
  for those due to the perturbative photon radation calculation, which is taken
  to be 100\% correlated between D0 and ATLAS, and to the
  photon energy cutoff, taken to be 100\% correlated between CDF and D0. 
  }  
\label{tbl:ewkelectrons}
\end{table}

\begin{table}[!tp]
  \centering 
  \begin{tabular}{lccc}
  \hline \hline 
    Uncertainty          & CDF       & ATLAS     & LHCb \\
    \hline 
    Perturbative photon rad.   & 2.3 (2.3) & 2.5 (3.5) & 8.6 \\
    Photon energy cutoff & 1 (1)  & ---          & --- \\ 
    FSR $e^+e^-$         & 1 (1)     & 0.8 (3.6) & --- \\ 
    Total                & 2.7 (2.7) & 2.6 (5.6) & 8.6 \\
\hline \hline
  \end{tabular}
  \caption{QED uncertainties in MeV on the $m_W$ measurement in the muon 
  channel for ATLAS and CDF using the \mT~(\pT) fit, and for LHCb. 
  The uncertainties are taken to be uncorrelated between the experiments.
  }  
\label{tbl:ewkmuons}
\end{table}

To estimate the uncertainty from the limitations of the shower model relative to the 
matrix-element calculation, D0 and ATLAS perform a direct comparison between PHOTOS 
and WGRAD~\cite{Baur:1998kt,Baur:2004ig} or WINHAC~\cite{Placzek:2003zg,Placzek:2009jy,Placzek:2013moa}, 
respectively.  Since ATLAS and D0 use the same shower model, their uncertainties are 
considered as correlated.  LHCb estimates the uncertainty with a hybrid approach of 
comparing \textsc{Powheg} with and without the NLO EW calculation, and the range of 
the PHOTOS, \textsc{Pythia8}, and \textsc{Herwig} shower models.  The average of the 
measurements from the different shower models is taken as the central value, so the 
uncertainty is considered as uncorrelated.  CDF uses a third strategy, applying a 
correction to the measurement using the 
HORACE~\cite{CarloniCalame:2003ux,CarloniCalame:2006zq,CarloniCalame:2007cd} 
generator, which matches multiple-photon radiation to the $O(\alpha)$ calculation.
The residual uncertainties are largely due to MC statistics, and are considered as 
uncorrelated.

The shower model includes a lower threshold on the emitted photon energy, expressed 
as a ratio with respect to the energy of the lepton from the $W$ boson decay.  
CDF uses a threshold of $10^{-5}$ and determines the uncertainty by increasing 
the threshold by a factor of 3.  D0 uses a similar procedure except with an increase 
from $2.5 \times 10^{-4}$ to $2 \times 10^{-2}$.  These uncertainties are taken to be 
completely correlated.

To account for the higher-order process of an off-shell final-state photon 
splitting into an $e^+ e^-$ pair, CDF applies an effective radiator 
approximation to the radiated photons.  ATLAS does not apply a correction, 
instead taking the uncertainty from a PHOTOS model of this process.  
The uncertainties are treated as uncorrelated.

%% file: procedures.tex
The combination of $m_W$ measurements is performed by first replicating each experiment's 
combination of fit results within the experiment, applying any relevant $\delta m_W$ shifts, 
and then combining across experiments.

The CDF individual $m_W$ values using the \mT, \pTl, and \met~distributions in the electron 
and muon channels are combined using the reported uncertainties and correlations, giving the
results shown in Table~\ref{tbl:cdfval}.  The CDF measurement applies $\delta m_W$ values of
$3.3$, $3.6$, and $3.0$~MeV, respectively, to fits to {\textsc{ResBos-C}} with the CTEQ6M PDF
set.  We remove these $\delta m_W$ corrections and add 
$\delta m_W^{\scriptsize\textrm{pol}} + \delta m_W^{\scriptsize\textrm{gen}} = -11.1$,
$-11.8$, and $-15.7$~MeV to the \mT, \pTl, and \met~results, respectively,
corresponding to the {\textsc{ResBos2}} calculation of leptonic angular distributions described
in Sec.~\ref{sec:polarization} and the removal of the $\ell\nu$ invariant mass requirement
discussed in Sec.~\ref{sec:resonance}.  Finally, a shift to the target PDF set is applied.
For the NNPDF3.1 PDF set this procedure
gives a combined CDF value of $m_W = 80432.1 \pm 9.4$~MeV, which is consistent with the published
CDF value of $m_W = 80433.5 \pm 9.4$~MeV within the uncertainty of the procedure.  

The principal component analysis used by CDF to reduce statistical effects in the PDF uncertainty
evaluation is not used in the combination with other experiments, since different measurements 
would give different principal components and complicate the correlation evaluations for the
combination.  Instead the Hessian sets provided by the NNPDF collaboration are used to estimate
the PDF uncertainty.  The combined CDF value with this uncertainty is labelled
``Combined ($\sigma_{\scriptsize\textrm{PDF}} = 6.6$~MeV)'' in Table~\ref{tbl:cdfval} and
corresponds to the entry labelled "NNPDF 3.1" in Table~\ref{tbl:tevatron_opt2} in
Sec.~\ref{sec:results}.

\begin{table*}[tbp]
  \centering
\begin{tabular*}{\linewidth}{l @{\extracolsep{\fill}}cccccc}
\hline \hline
                & Published $m_W$ & Input $m_W$ & $\delta m_W^{\scriptsize\textrm{pol}}$ & $\delta m_W^{\scriptsize\textrm{gen}}$
                & $\delta m_W^{\scriptsize\textrm{PDF}}$ & LHC-TeV MWWG $m_W$ 
                \\
\hline
\mT($e,\nu$)    & 80429.1     & 80425.8     & $-9.5$     & $-1.6$       &  $13.1$          & 80427.8 \\
\pTl($e$)       & 80411.4     & 80407.8     & $-8.4$     & $-3.4$       &  $14.0$          & 80410.0 \\
\met($e$)       & 80426.3     & 80423.3     & $-12.5$    & $-3.2$       &  $15.1$          & 80422.7 \\
\mT($\mu,\nu$)  & 80446.1     & 80442.8     & $-9.5$     & $-1.6$       &  $13.1$          & 80444.8 \\
\pTl($\mu$)     & 80428.2     & 80424.6     & $-8.4$     & $-3.4$       &  $14.0$          & 80426.8 \\
\met($\mu$)     & 80428.9     & 80425.9     & $-12.5$    & $-3.2$       &  $15.1$          & 80425.3 \\
\hline
Combined ($\sigma_{\scriptsize\textrm{PDF}} = 3.9$~MeV)  & 80433.5  & &   &                &          & 80432.1 \\
Combined ($\sigma_{\scriptsize\textrm{PDF}} = 6.6$~MeV)  &          & &   &                &          & 80433.3 \\
\hline \hline
\end{tabular*}
 \caption{Published CDF values and input values to the combination, where the latter correspond to the results 
 obtained from the direct CDF fits to {\textsc{ResBos-C}} with the CTEQ6M PDF set.  The combination procedure applies
 shifts to these results to update to the {\textsc{ResBos2}} calculation
 ($\delta m_W^{\scriptsize\textrm{pol}}$ and $\delta m_W^{\scriptsize\textrm{gen}}$)
 and a shift to update to the NNPDF3.1 PDF set ($\delta m_W^{\scriptsize\textrm{PDF}}$).  The total statistical and
 systematic uncertainties on the shifts are 1.2, 1.1, and 2.1 MeV for the \mT, \pTl, and \met~fits respectively.
 The combined value
 is consistent with that obtained by CDF when using the PDF uncertainties determined by CDF, labelled
 ``Combined ($\sigma_{\scriptsize\textrm{PDF}} = 3.9$~MeV)''.  When combining the result with other measurements, the
 uncertainty is evaluated using NNPDF3.1 eigenvectors to give the result labelled
 ``Combined ($\sigma_{\scriptsize\textrm{PDF}} = 6.6$~MeV)''.  The difference is due to a change in the weight of
 each fit distribution.  All units are in MeV.
}
\label{tbl:cdfval}
\end{table*}

The individual D0 measurements with the \mT~and \pTl~distributions using data sets corresponding 
to 1.1~fb$^{-1}$ and 4.3~fb$^{-1}$ are combined using the reported uncertainties to give the result 
$m_W = 80375.1 \pm 23.1$~MeV, which is the value quoted by D0 rounded to the nearest MeV.  Before 
combining with other measurements, a number of shifts are applied.  First, a shift from CTEQ6.1 to
CTEQ6.6 is applied to the measurement based on 1.1~fb$^{-1}$ of integrated luminosity.  Shifts of 
$\delta m_W^{\scriptsize\textrm{pol}} = -6.4$, $-6.9$, and $-15.8$ are applied to the \mT, \pTl,
and \met fit results, respectively, to update the {\textsc{ResBos-CP}} leptonic angular distributions to
those of {\textsc{ResBos2}}.  Finally, a $\delta m_W^{\mathrm\Gamma}$ shift adjusts ${\mathrm\Gamma_W}$
to that of the SM prediction.  The result with these shifts and the published D0 PDF uncertainty
of $\approx 11$~MeV is labelled ``Combined ($\sigma_{\scriptsize\textrm{PDF}} = 11$~MeV)'' in
Table~\ref{tbl:d0val}.  The value with uncertainties updated to those calculated with {\textsc{Wj-MiNNLO}}
and CTEQ6.6 is $m_W = 80377.9 \pm 25.5$~MeV and is labelled ``Combined ($\sigma_{\scriptsize\textrm{PDF}} = 15.1$~MeV)''
in the table.

\begin{table*}[tbp]
  \centering
\begin{tabular*}{\linewidth}{l @{\extracolsep{\fill}}ccccc}
\hline \hline 
  & Published $m_W$   & $\delta m_W^{\scriptsize\textrm{pol}}$ & $\delta m_W^{\scriptsize\textrm{PDF}}$
  & $\delta m_W^{\mathrm\Gamma}$ & LHC-TeV MWWG $m_W$ \\
\hline
Run 2a \mT($e,\nu$)  & 80401          & $-6.4$           & $14.3$         &  1.4 & 80410.3      \\
Run 2a \pTl($e$)     & 80400          & $-6.9$           & $16.7$         &  1.4 & 80411.2      \\
Run 2a \met($e$)     & 80402          & $-15.8$          & $17.5$         &  1.4 & 80405.1      \\
Run 2b \mT($e,\nu$)  & 80371          & $-6.4$           &                &  1.5 & 80366.1      \\
Run 2b \pTl($e$)     & 80343          & $-6.9$           &                &  1.5 & 80337.6     \\
\hline
Combined ($\sigma_{\scriptsize\textrm{PDF}} = 11$~MeV)           & 80375      &         & &      & 80373.4      \\ 
Combined ($\sigma_{\scriptsize\textrm{PDF}} = 15.1$~MeV)         &            &         & &      & 80377.9      \\ 
\hline \hline
\end{tabular*}
  \caption{Published D0 values corresponding to the CTEQ6M (Run 2a) and CTEQ6.6 (Run 2b) PDF sets, 
along with the following shifts: modifying the leptonic angular distributions to match those of
{\textsc{ResBos2}} ($\delta m_W^{\scriptsize\textrm{pol}}$);
modifying the Run 2a result to correspond to the CTEQ6.6 PDF set ($\delta m_W^{\scriptsize\textrm{PDF}}$); 
and modifying the $W$ boson width to the Standard Model prediction ($\delta m_W^{\mathrm\Gamma}$).
 The total statistical and systematic uncertainties on the shifts are 1.2, 1.2, and 2.3 MeV for the
 \mT, \pTl, and \met~fits respectively.
The combined result with the published D0 PDF uncertainty obtained using {\textsc Pythia} and the CTEQ6.1
PDF set is labelled ``Combined ($\sigma_{\scriptsize\textrm{PDF}} = 11$~MeV)'', and the result with
PDF uncertainties updated to those of CTEQ6.6 calculated with {\textsc{Wj-MiNNLO}} is labelled
``Combined ($\sigma_{\scriptsize\textrm{PDF}} = 15.1$~MeV)''.  The results differ due to different
weights of the individual fits to kinematic distributions.  All units are in MeV.  }
\label{tbl:d0val}
\end{table*}

The ATLAS measurement is reproduced using the parameterized simulation to give a value of 
$m_W = 80369.7 \pm 18.5$~MeV, which is within a few tenths of an MeV of the published result.  
A $\delta m_W^{\mathrm\Gamma} = 0.7$~MeV correction is added to update ${\mathrm\Gamma}_W$, and further
$\delta m_W^{\scriptsize\textrm{PDF}}$ shifts are applied to provide the central value for the
target PDF set.  All LHCb $m_W$ values for 
the combination are determined from a direct fit to the data using the target PDF set, so only a 
$\delta m_W^{\mathrm\Gamma} = -0.7$~MeV shift is applied to update the value of the $W$ boson width.

%% file: results.tex
A series of combinations are performed corresponding to the Tevatron Run 2 experiments, the LHC experiments, 
all experiments including the result from the LEP combination, and all experiments except one.  For each 
experiment the central value, uncertainty, and $\chi^2$ of the individual measurements is shown for 
the ABMP16, CT14, CT18, MMHT2014, MSHT20, NNPDF31, and NNPDF40 PDF sets.  For the combined result of 
multiple experiments the overall PDF uncertainty is also shown.  The PDF uncertainties for the individual 
experiments are given in Table~\ref{tab:pdfunc}.


\subsubsection{Hadron-collider measurements}

Results for the Tevatron Run 2 experiments are listed in Table~\ref{tbl:tevatron_opt2}. The individual 
combinations of the CDF and D0 fit results are satisfactory for all PDF sets, with probabilities ranging
from 12\% to 24\%.  The Tevatron-wide combination has a total uncertainty ranging from 8.9~MeV for ABMP16
to 15.9~MeV for CT18, and a $\chi^2$ probability of 0.5--0.8\%.

As discussed in Section~\ref{sec:corrPDF}, PDF uncertainties are fully correlated between CDF and D0. 
The PDF uncertainty in the combination is therefore close to that obtained for each experiment, and 
ranges from 4~MeV for ABMP16 to 13.5~MeV for CT18. The combined central value ranges from 80408.2~MeV 
for ABMP16 to 80433.4~MeV for NNPDF4.0.  The difference between the NNPDF3.1 and NNPDF4.0 combinations, 
8.4~MeV, is similar to the PDF uncertainty of the NNPDF4.0 set (7.8 MeV).  Similar trends are observed 
for the CDF and D0 measurements separately.  Some variation in the results with PDF set is expected due
to differences in input data sets to the PDFs, and to the differences in the compatibility with Drell-Yan
measurements discussed in Section~\ref{sec:PDFWZ}.  Further understanding of these differences would benefit
future combinations.

\begin{table*}[!tp]
  \centering
  \input{tevatron_evolution_precomb.tex}
  \caption{The CDF and D0 Run 2 $m_W$ and $\chi^2$ values obtained from a combination of the individual
  measurement distributions and decay channels, along with the combined Tevatron Run 2 $m_W$, PDF uncertainty,
  $\chi^2$, and probability of obtaining this $\chi^2$ or larger.  Mass units are in MeV.
}
\label{tbl:tevatron_opt2}
\end{table*}


The LHC results are summarized in Table~\ref{tbl:lhc_opt2}. The $\chi^2$ per degree of freedom of the ATLAS
combination ranges from 29/27 (for NNPDF3.1) to 45/27 (for MSHT20).  The latter corresponds to a probability
of about 2\%.  The larger $\chi^2$ for MSHT20 is consistent with the calculations of Drell-Yan measurements. 
The ATLAS and LHCb measurements are compatible and have a total
uncertainty ranging from 14.2~MeV to 16.6~MeV.

\begin{table*}[!tp]
  \centering
  \input{lhc_evolution_precomb.tex}
  \caption{The ATLAS and LHCb $m_W$ values obtained from a combination of the individual measurement distributions
  and decay channels, along with the combined LHC $m_W$, PDF uncertainty, and $\chi^2$, and probability of obtaining
  this $\chi^2$ or larger.  The $\chi^2$ 
  of the combination of fit distributions and decay channels is shown for ATLAS; no $\chi^2$ is shown for LHCb as the
  measurement is performed using one distribution in one channel.  Mass units are in MeV.
  }
  \label{tbl:lhc_opt2}
\end{table*}

\begin{figure}[!tp]
  \centering
  \includegraphics[width=0.95\columnwidth]{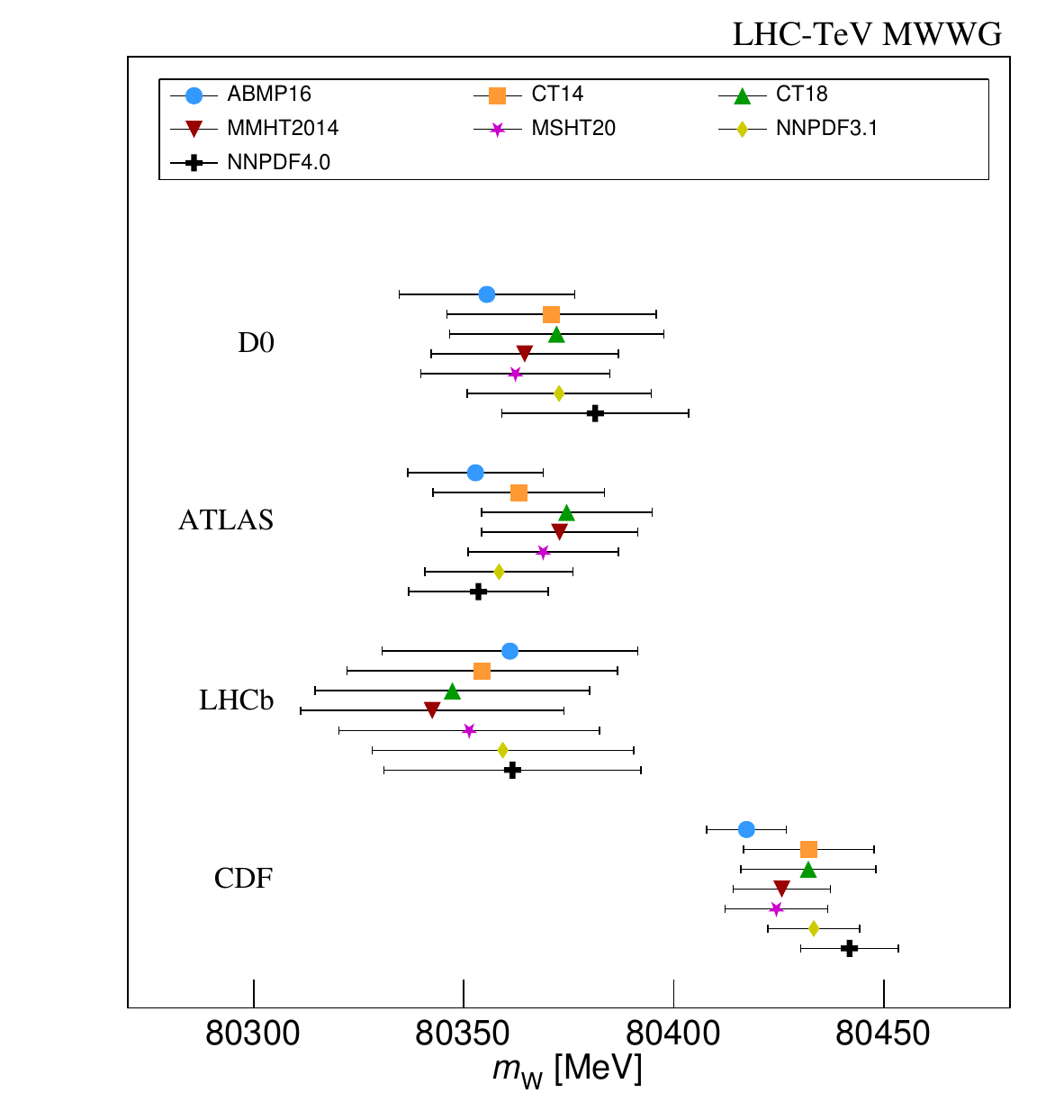}
  \caption{The D0, ATLAS, LHCb, and CDF $m_W$ values and uncertainties using the ABMP16, CT14, CT18, MMHT2014,
  MSHT20, NNPDF3.1, and NNPDF4.0 PDF sets. 
}
\label{fig:sumexpt}
\end{figure}

The individual experimental results are shown in Figure~\ref{fig:sumexpt} for all considered PDF sets.
The combination of ATLAS and LHCb measurements benefits from anti-correlated PDF uncertainties~\cite{Bozzi:2015zja}. 
Therefore the combined PDF uncertainties and the variation of the combined central values are smaller than for the 
individual experiments. The ATLAS $m_W$ value ranges from 80352.8~MeV for ABMP16 to 80374.5~MeV for CT18. This range
is comparable to that of the Tevatron experiments. A similar spread but opposite trends are observed for LHCb, and
the spread of $m_W$ values is reduced from $\approx 20$~MeV to 14.1~MeV in the combination. The PDF uncertainties
range from 4.0~MeV to 11.4~MeV for ATLAS and 3.0 to 12.2~MeV for LHCb, but are reduced to 2.9--6.5~MeV for
the combined result.  

\subsubsection{All measurements}

Tables~\ref{tbl:all1_opt2}--\ref{tbl:all3_opt2} provide the results for various combinations including
LEP, whose uncertainties are treated as uncorrelated with the others.  A combination of all measurements
yields a total uncertainty ranging between 7.5 and 11.5~MeV, though the $\chi^2$ probabilities are low,
ranging from $8 \times 10^{-6}$ to $5\times 10^{-3}$.  The low probabilities
reflect the discrepancy between the CDF measurement and the other measurements. The combined value of
$m_W$ for the CT18 PDF set, which gives the largest compatibility with the broader Drell-Yan measurements,
is $m_W = 80394.6 \pm 11.5$~MeV with a probability of 0.5\%.  The relative weights of the CDF, ATLAS, LHCb,
LEP, and D0 measurements are 41\%, 28\%, 13\%, 12\%, and 5\%, respectively.  Weights for other PDF sets are
given in the Appendix.  The largest difference in $m_W$ between PDF sets is 10.4 MeV.

  \input{world_evolution1_precomb.tex}

\input{world_evolution2_precomb.tex}

\input{world_evolution3_precomb.tex}

A possible procedure for combining measurements with low compatibility is to scale all uncertainties by the
square root of the ratio of the $\chi^2$ to the number of degrees of freedom.
This procedure effectively assumes a common underestimated uncertainty, which is an unlikely scenario for
these measurements.  The PDF uncertainty is only partially correlated, and the uncertainty from the CT18 set
is the most conservative.  Other measurement uncertainties are smaller or are statistically constrained and
therefore uncorrelated.  Further measurements or studies are required to obtain more consistent results.

To evaluate the significance of differences between individual measurements and the others, separate
combinations are performed excluding, in turn, each individual result from the average.
Removing LEP, D0, or LHCb from the combination increases the uncertainty by up to 0.9~MeV and affects
the central value by up to 8~MeV.
When removing ATLAS the
$\chi^2$ probability ranges from 0.3\% to 1.2\%, and the uncertainty ranges from 8.3~to 13.2~MeV.
The
combinations with CDF excluded have good compatibility and the total uncertainty increases to 11.2--13.3~MeV,
or 2--4~MeV more than the full combination.  The variation of this combination with PDF set is 11.9 MeV, with 
the value for the ABMP16 PDF set considerably lower than the others (the variation is 4.5 MeV without this set).
The combination of all measurements except CDF is $m_W=80369.2 \pm 13.3$~MeV for the CT18 PDF set, with a
91\% probability of consistency.  The relative weights for the ATLAS, D0, LHCb, and LEP measurements are 42\%,
23\%, 18\%, and 16\%, respectively.
 
The partial combinations are also used to evaluate the difference between each experimental result and the
combination of the others. Considering all PDF sets, the LEP result is compatible with the average of the
others to better than one standard deviation. The compatibility of D0 or LHCb with the rest ranges from
1--1.8 standard deviations. The ATLAS result differs from the others by 1.6--3.6 standard deviations,
where the largest difference is obtained with the NNPDF4.0 PDF set. Finally, the CDF measurement differs
from the others by 3.6--5.2 standard deviations, depending on the choice of the PDF set. The smallest 
significance corresponds to the CT18 set and the largest significance corresponds to the NNPDF4.0 set.
 
\begin{figure*}[p]
  \centering
  \includegraphics[width=0.95\columnwidth]{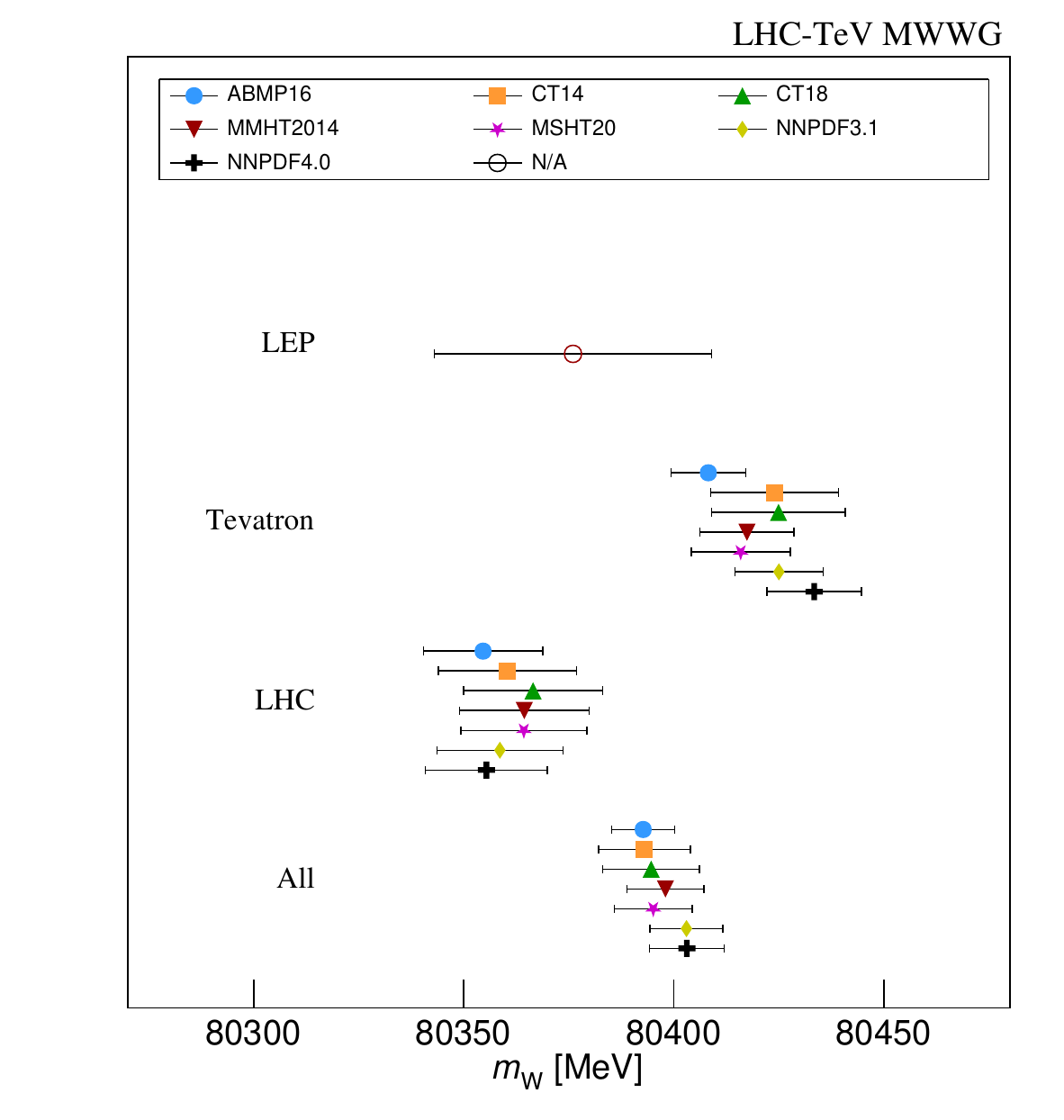}
  \includegraphics[width=0.95\columnwidth]{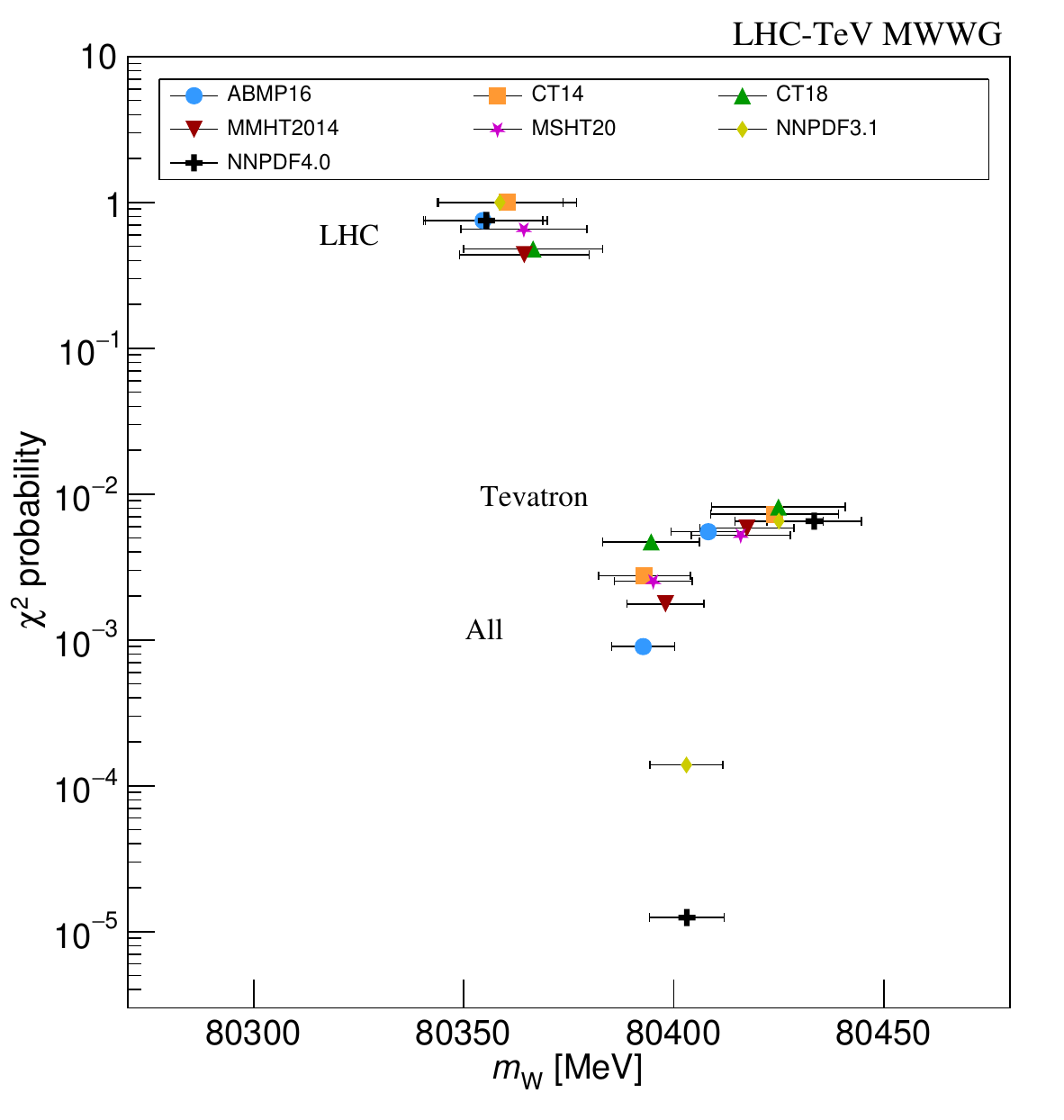}
  \includegraphics[width=0.95\columnwidth]{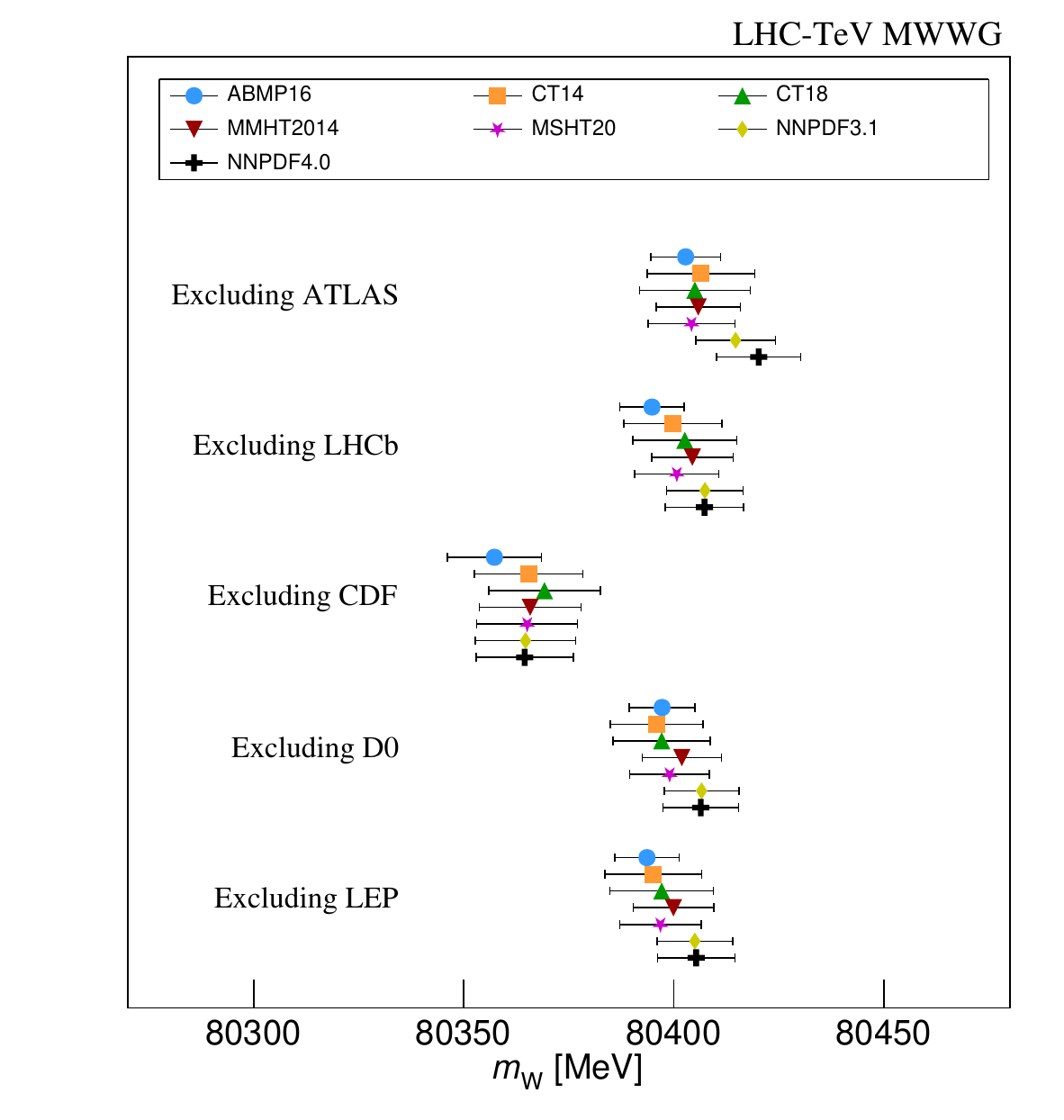}
  \includegraphics[width=0.95\columnwidth]{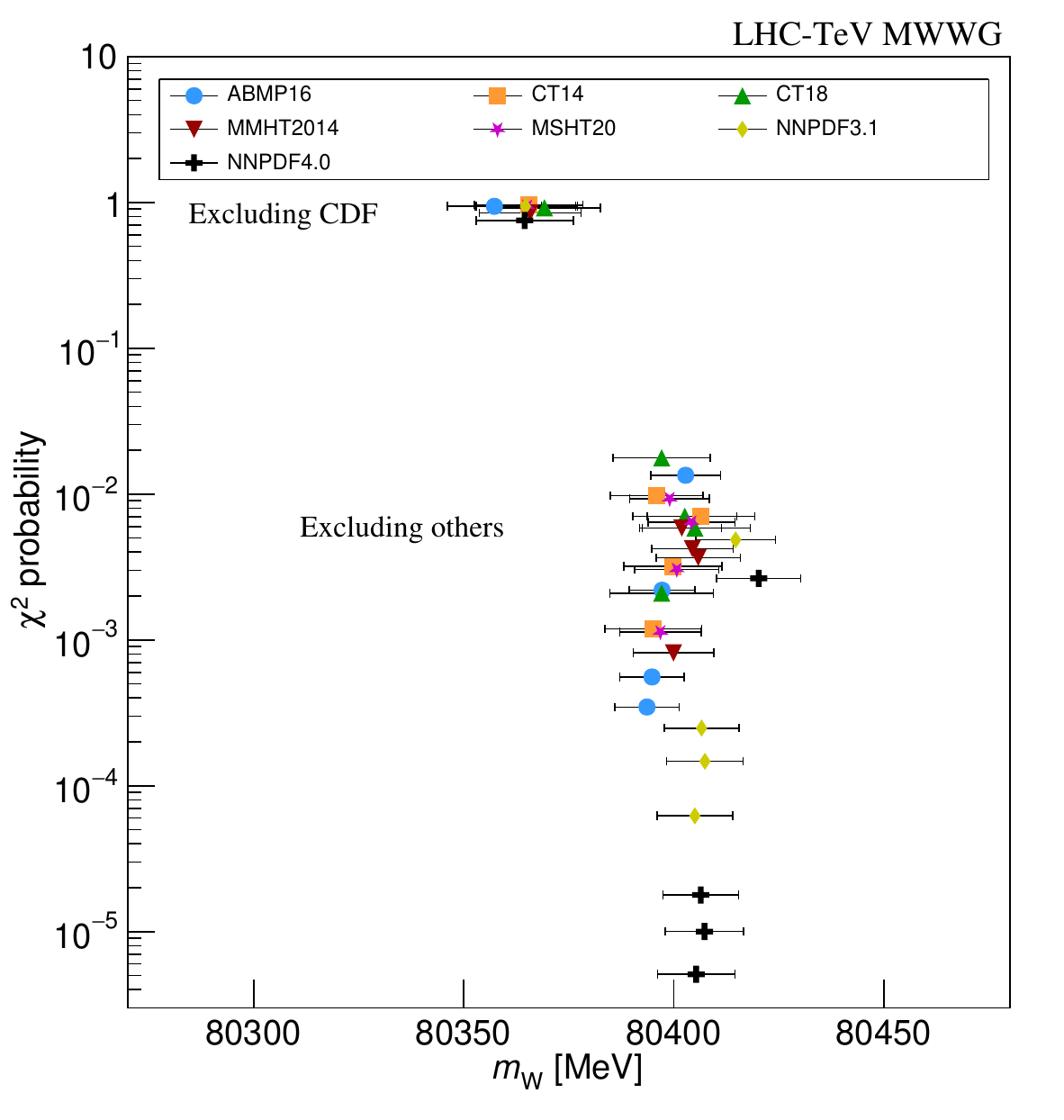}
  \caption{Top left: The combined $m_W$ values and uncertainties from LEP, the Tevatron, LHC, and all experiments,
  using the ABMP16, CT14, CT18, MMHT2014, MMHT20, NNPDF3.1, and NNPDF4.0 PDF sets.  Right: The corresponding
  probability of consistency determined using the $\chi^2$ per degrees of freedom.
 Bottom left: The combined $m_W$ values and uncertainties for all experiments except one using the ABMP16,
  CT14, CT18, MMHT2014, MMHT20, NNPDF3.1, and NNPDF4.0 PDF sets.  Right: The corresponding probability of
  consistency determined using the $\chi^2$ per degrees of freedom.
}
  \label{fig:sumcomb}
\end{figure*}

The $m_W$ combinations from LEP, the Tevatron, the LHC, and all experiments are presented in Figure~\ref{fig:sumcomb}
for all PDF sets, along with the corresponding $\chi^2$ probabilities. The same information is also shown
for the combinations removing one experimental result at a time.

%% file: tevatron_evolution_precomb.tex
\begin{tabular*}{\linewidth}{l @{\extracolsep{\fill}}llllllllll}
\hline \hline
   & \multicolumn{2}{c}{CDF (5 d.o.f.)} & & \multicolumn{2}{c}{D0 (4 d.o.f.)} & & \multicolumn{3}{c}{Tevatron Run 2 (1 d.o.f.)} \\
PDF set   & \multicolumn{1}{c}{$m_W$}   & $\chi^2$ & & \multicolumn{1}{c}{$m_W$} & $\chi^2$ & & \multicolumn{1}{c}{$m_W$} & $\sigma_{\scriptsize\textrm{PDF}}$ & $\chi^2$ & p($\chi^2,n$) \\
\hline
   ABMP16 & $80417.3~\pm~9.5$  &  8.8  & & $80355.4 \pm 20.9$ &  6.6   & & $80408.2 \pm  8.9$ &   4.0 &      7.7  & 0.6\% \\
     CT14 & $80432.1 \pm 15.5$ &  7.7  & & $80370.9 \pm 24.9$ &  5.9   & & $80424.0 \pm 15.2$ &  12.6 &      7.2  & 0.7\% \\
     CT18 & $80432.0 \pm 16.1$ &  7.6  & & $80372.0 \pm 25.5$ &  5.9   & & $80424.9 \pm 15.9$ &  13.5 &      7.0  & 0.8\% \\
 MMHT2014 & $80425.7 \pm 11.6$ &  7.0  & & $80364.4 \pm 22.3$ &  5.5   & & $80417.4 \pm 11.2$ &   7.8 &      7.6  & 0.6\% \\
   MSHT20 & $80424.4 \pm 12.2$ &  7.6  & & $80362.3 \pm 22.5$ &  6.1   & & $80415.9 \pm 11.8$ &   8.6 &      7.8  & 0.5\% \\
 NNPDF3.1 & $80433.3 \pm 10.9$ &  7.6  & & $80372.7 \pm 21.9$ &  5.8   & & $80425.0 \pm 10.5$ &   6.8 &      7.4  & 0.7\% \\
 NNPDF4.0 & $80441.8 \pm 11.6$ &  7.2  & & $80381.3 \pm 22.2$ &  5.7   & & $80433.4 \pm 11.2$ &   7.8 &      7.4  & 0.7\% \\
\hline \hline 
\end{tabular*}

%% file: lhc_evolution_precomb.tex
\begin{tabular*}{\linewidth}{l @{\extracolsep{\fill}}llllllllll}
\hline \hline
           &   \multicolumn{2}{c}{ATLAS (27 d.o.f)} & & \multicolumn{2}{c}{LHCb} & & \multicolumn{3}{c}{LHC (1 d.o.f)} \\
PDF set    & \multicolumn{1}{c}{$m_W$}  & $\chi^2$ & & \multicolumn{1}{c}{$m_W$} & $\chi^2$ & & \multicolumn{1}{c}{$m_W$} & $\sigma_{\scriptsize\textrm{PDF}}$ & $\chi^2$ & p($\chi^2,n$) \\
\hline 
    ABMP16 & $80352.8 \pm 16.1$ & 31 & & $80361.0 \pm 30.4$ & --  & & $80354.6 \pm 14.2$ &    2.9 &    0.1 & 75\% \\
      CT14 & $80363.1 \pm 20.4$ & 30 & & $80354.4 \pm 32.2$ & --  & & $80360.4 \pm 16.4$ &    6.5 &    0.0 & 100\% \\
      CT18 & $80374.5 \pm 20.3$ & 30 & & $80347.3 \pm 32.7$ & --  & & $80366.5 \pm 16.6$ &    6.3 &    0.5 & 48\% \\
  MMHT2014 & $80372.8 \pm 18.6$ & 30 & & $80342.5 \pm 31.3$ & --  & & $80364.4 \pm 15.4$ &    5.1 &    0.6 & 44\% \\
    MSHT20 & $80368.9 \pm 17.9$ & 45 & & $80351.3 \pm 31.0$ & --  & & $80364.3 \pm 15.0$ &    4.5 &    0.2 & 65\% \\
  NNPDF3.1 & $80358.4 \pm 17.6$ & 29 & & $80359.3 \pm 31.1$ & --  & & $80358.6 \pm 15.0$ &    5.0 &    0.0 & 100\% \\
  NNPDF4.0 & $80353.5 \pm 16.6$ & 35 & & $80361.6 \pm 30.6$ & --  & & $80355.4 \pm 14.5$ &    3.8 &    0.1 & 75\% \\
\hline \hline
\end{tabular*}

%% file: world_evolution1_precomb.tex
\begin{table}[!tp]
  \centering
\begin{tabular}{lllll}
\hline \hline
                                     \multicolumn{5}{c}{All experiments (4 d.o.f.)} \\
PDF set                      & \multicolumn{1}{c}{$m_W$} & $\sigma_{\scriptsize\textrm{PDF}}$ & $\chi^2$ & p($\chi^2,n$) \\
\hline
                             ABMP16 & $80392.7 \pm  7.5$ &   3.2 &    29  & 0.0008\% \\
                               CT14 & $80393.0 \pm 10.9$ &   7.1 &    16  & 0.3\% \\
                               CT18 & $80394.6 \pm 11.5$ &   7.7 &    15  & 0.5\% \\
                           MMHT2014 & $80398.0 \pm  9.2$ &   5.8 &    17  & 0.2\% \\
                             MSHT20 & $80395.1 \pm  9.3$ &   5.8 &    16  & 0.3\% \\
                           NNPDF3.1 & $80403.0 \pm  8.7$ &   5.3 &    23   & 0.1\% \\
                           NNPDF4.0 & $80403.1 \pm  8.9$ &   5.3 &    28   & 0.001\% \\
\hline \hline
\end{tabular}
\caption{Combination of $m_W$ measurements from the individual experiments.
  Shown for each PDF are the PDF uncertainty, $\chi^2$, and probability of
  obtaining this $\chi^2$ or larger.  Mass units are in MeV.
  \label{tbl:all1_opt2}}
\end{table}

\begin{table}[!tp]
  \centering
\begin{tabular}{lllll}
\hline \hline

                                     \multicolumn{5}{c}{All except LEP (3 d.o.f.)} \\
PDF set                      & \multicolumn{1}{c}{$m_W$} & $\sigma_{\scriptsize\textrm{PDF}}$ & $\chi^2$ & p($\chi^2,n$) \\
\hline
                             ABMP16 & $80393.6 \pm  7.8$ &     3.4 &     19   & 0.03\% \\
                               CT14 & $80395.1 \pm 11.6$ &     8.0 &     16   & 0.1\% \\
                               CT18 & $80397.1 \pm 12.3$ &     8.8 &     15   & 0.2\% \\
                           MMHT2014 & $80399.9 \pm  9.6$ &     6.2 &     17   & 0.7\% \\
                             MSHT20 & $80396.8 \pm  9.7$ &     6.3 &     16   & 0.1\% \\
                           NNPDF3.1 & $80405.0 \pm  9.0$ &     5.6 &     22   & 0.007\% \\
                           NNPDF4.0 & $80405.3 \pm  9.2$ &     5.7 &     27   & 0.0006\% \\
\hline \hline
\end{tabular}
\caption{Combination of all $m_W$ measurements except the LEP average.  Shown for each PDF are the 
  PDF uncertainty, $\chi^2$, and probability of
  obtaining this $\chi^2$ or larger.  Mass units are in MeV.
  \label{tbl:all2_opt2}
}
\end{table}

%% file: world_evolution2_precomb.tex
\begin{table}[!tp]
  \centering
\begin{tabular}{lllll}
\hline \hline 
                                       \multicolumn{5}{c}{All except CDF (3 d.o.f.)} \\
PDF set                      & \multicolumn{1}{c}{$m_W$} & $\sigma_{\scriptsize\textrm{PDF}}$ & $\chi^2$ & p($\chi^2,n$) \\
\hline
                             ABMP16 & $80357.3 \pm 11.2$ &  2.6 &  0.4 & 94\% \\
                               CT14 & $80365.4 \pm 12.9$ &  5.8 &  0.3 & 96\% \\
                               CT18 & $80369.2 \pm 13.3$ &  6.2 &  0.5 & 92\% \\
                           MMHT2014 & $80365.8 \pm 12.1$ &  4.7 &  0.8 & 85\% \\
                             MSHT20 & $80365.1 \pm 12.0$ &  4.4 &  0.4 & 94\% \\
                           NNPDF3.1 & $80364.7 \pm 11.9$ &  4.5 &  0.4 & 94\% \\
                           NNPDF4.0 & $80364.5 \pm 11.6$ &  3.9 &  1.2 & 75\% \\
\hline \hline
\end{tabular}
\caption{Combination of $m_W$ measurements from all individual experiments except CDF.
  Shown for each PDF are the PDF uncertainty, $\chi^2$, and probability of
  obtaining this $\chi^2$ or larger.  Mass units are in MeV.
  \label{tbl:all3_opt2}}
\end{table}

\begin{table}[!tp]
  \centering
\begin{tabular}{lllll}
\hline \hline 
                                    \multicolumn{5}{c}{All except D0 (3 d.o.f.)} \\
PDF set                      & \multicolumn{1}{c}{$m_W$} & $\sigma_{\scriptsize\textrm{PDF}}$ & $\chi^2$ & p($\chi^2,n$)\\
\hline
                             ABMP16 & $80397.2 \pm  7.8$ &  3.1 &    15 & 0.2\% \\
                               CT14 & $80395.9 \pm 11.0$ &  7.0 &    11 & 1.2\% \\
                               CT18 & $80397.1 \pm 11.6$ &  7.6 &    10 & 1.9\% \\
                           MMHT2014 & $80401.9 \pm  9.4$ &  5.6 &    13 & 0.5\% \\
                             MSHT20 & $80399.0 \pm  9.5$ &  5.7 &    12 & 0.7\% \\
                           NNPDF3.1 & $80406.6 \pm  8.9$ &  5.1 &    19 & 0.03\% \\
                           NNPDF4.0 & $80406.4 \pm  9.0$ &  5.1 &    25 & 0.002\% \\
\hline \hline
\end{tabular}
\caption{Combination of $m_W$ measurements from all individual experiments except D0.
  Shown for each PDF are the PDF uncertainty, $\chi^2$, and probability of
  obtaining this $\chi^2$ or larger.  Mass units are in MeV.
  \label{tbl:all4_opt2}}
\end{table}

%% file: world_evolution3_precomb.tex
\begin{table}[!tp]
  \centering
  \begin{tabular}{lllll}
\hline \hline
                              \multicolumn{5}{c}{All except ATLAS (3 d.o.f.)} \\
PDF set                      & \multicolumn{1}{c}{$m_W$} & $\sigma_{\scriptsize\textrm{PDF}}$ & $\chi^2$ & p($\chi^2,n$)\\
\hline 
                             ABMP16 & $80402.8 \pm  8.3$ &  3.5 & 11 & 1.2\% \\
                               CT14 & $80406.5 \pm 12.8$ &  9.1 & 12 & 0.7\% \\
                               CT18 & $80405.0 \pm 13.2$ &  9.6 & 13 & 0.5\% \\
                           MMHT2014 & $80405.8 \pm 10.0$ &  6.3 & 14 & 0.3\% \\
                             MSHT20 & $80404.2 \pm 10.3$ &  6.6 & 12 & 0.7\%  \\
                           NNPDF3.1 & $80414.7 \pm  9.5$ &  5.6 & 13 & 0.5\%  \\
                           NNPDF4.0 & $80420.2 \pm 10.0$ &  6.2 & 14 & 0.3\% \\
                            \hline \hline 
\end{tabular}
  \caption{Combination of $m_W$ measurements from the individual experiments except for ATLAS.
      Shown for each PDF are the PDF uncertainty, $\chi^2$, and probability of
  obtaining this $\chi^2$ or larger.  Mass units are in MeV.
    \label{tbl:all5_opt2}}
\end{table}

\begin{table}[!tp]
  \centering  
\begin{tabular}{lllll}
\hline \hline
                            \multicolumn{5}{c}{All except LHCb (3 d.o.f.)} \\
PDF set                   & \multicolumn{1}{c}{$m_W$} & $\sigma_{\scriptsize\textrm{PDF}}$ & $\chi^2$ & p($\chi^2,n$) \\
\hline 
                          ABMP16 & $80394.8 \pm  7.7$ & 3.4 &   18 & 0.04\%  \\
                            CT14 & $80399.8 \pm 11.7$ & 8.4 &   14 & 0.3\% \\
                            CT18 & $80402.6 \pm 12.4$ & 9.0 &   12 & 0.7\% \\
                        MMHT2014 & $80404.4 \pm  9.7$ & 6.5 &   13 & 0.5\% \\
                          MSHT20 & $80400.7 \pm 10.0$ & 6.8 &   14 & 0.3\% \\
                        NNPDF3.1 & $80407.4 \pm  9.1$ & 5.8 &   20 & 0.02\% \\
                        NNPDF4.0 & $80407.3 \pm  9.3$ & 5.9 &   26 & 0.001\% \\
\hline \hline 
\end{tabular}
  \caption{Combination of $m_W$ measurements from the individual experiments except for LHCb.
          Shown for each PDF are the PDF uncertainty, $\chi^2$, and probability of
  obtaining this $\chi^2$ or larger.  Mass units are in MeV.
    \label{tbl:all6_opt2}}
\end{table}

%% file: conclusion.tex
A combination of $m_W$ measurements from the CDF, D0, ATLAS, LHCb, and combined LEP experiments has been performed.
Where necessary, measurement results have been updated to incorporate an improved theoretical description of the 
final state distributions. Experimental resolution effects, which are required to propagate the impact of variations
in the theoretical description of $W$-boson production and decay, are accounted for using a realistic emulation of
the ATLAS, CDF, and D0 measurement procedures.  Results for LHCb are produced using the published analysis procedures.
 
The largest theoretical uncertainty arises from the parton distribution functions.  Results are presented for the two
most recent PDF sets from the NNPDF, CTEQ, and M(M/S)HT collaborations, as well as the most recent set from the ABMP
collaboration.  Partial or negative correlations of PDF uncertainties between the Tevatron, ATLAS, and LHCb experiments
reduce the dependence of the combined result on the PDF set.  This dependence is nonetheless significant, as the differences
between individual sets is of the same order as the associated uncertainty.  The dependence
of the measurements on PDF set are due to differences in the input data sets and to the modelling assumptions in the PDFs,
and could ultimately limit the precision of future $m_W$ measurements and combinations. Improving the experimental
precision on $m_W$ requires a better understanding of PDF model dependence, and of uncertainty correlations between PDF
sets.

The consistency of Drell-Yan cross-section measurements, as well as the $m_W$ combination, is highest for the CT18 PDF
set due to its large uncertainties.  With this PDF set the combination of LEP, LHC, and Tevatron Run 2 measurements
gives a value $m_W = 80394.6 \pm 11.5$~MeV.  This value has a $\chi^2$ probability of $0.5\%$ and is therefore disfavoured.
Other PDF sets give probabilities of consistency between $2\times 10^{-5}$ and $3\times 10^{-3}$.

Good consistency is observed when all experiments other than CDF are combined, with a resulting $W$-boson mass of
$80369.2 \pm 13.3$~MeV and a 91\% probability of consistency for the CT18 PDF set.  When using this set and uncertainty
for the CDF measurement and for the combination of the others, the values differ by 3.6 standard deviations.  Further
measurements or studies of procedures and uncertainties are required to improve the understanding and consistency of a 
world-average value of the $W$ boson mass.

%% file: appendix.tex
\begin{figure}[!tbp]
   \centering
        \includegraphics[width=0.9\columnwidth]{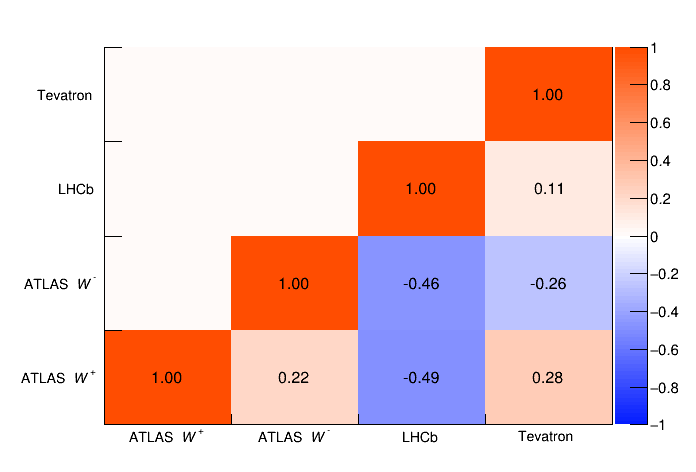}
        \includegraphics[width=0.9\columnwidth]{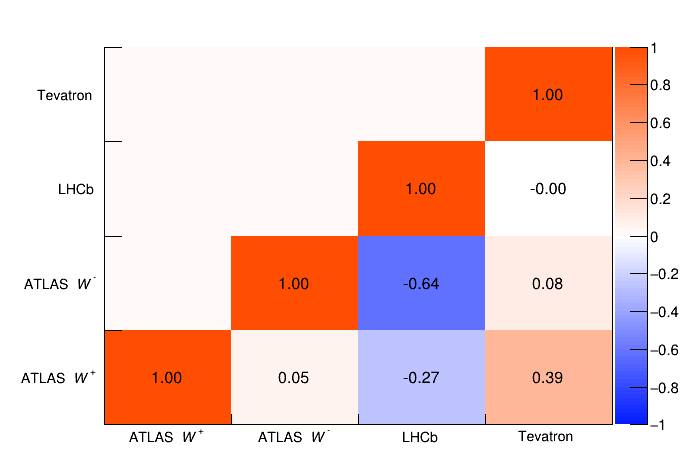}
\includegraphics[width=0.9\columnwidth]{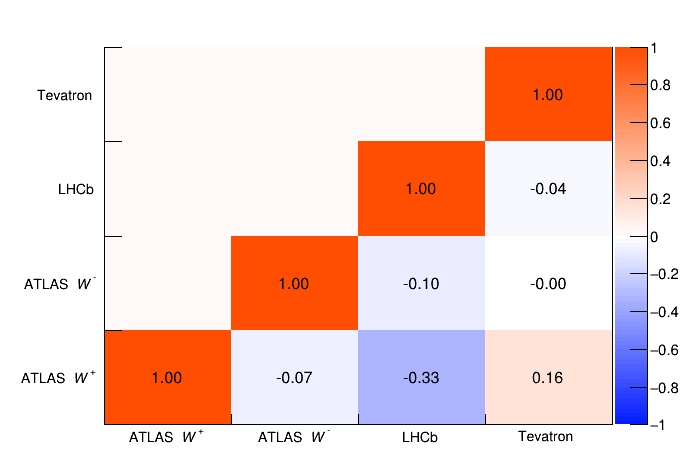}
\caption{PDF uncertainty correlation matrices for the CT14, MMHT2014, and NNPDF3.1 PDF sets, shown from top to bottom.
}
\label{fig:pdfcorr2}
\end{figure}

We provide here additional information on the combination inputs and results.
Figure~\ref{fig:pdfcorr2} shows the correlation matrices for the hadron-collider measurements 
for the CT14, MMHT2014, NNPDF3.1 PDF sets.  Tables~\ref{tbl:wgttev}--\ref{tbl:wgtall} give
the relative weight of each measurement to various combinations of measurements.

\begin{table*}[!tbp]
  \centering
  \input{TeVatron_weight_table_precomb.tex}
  \caption{Relative weights (in percent) of the CDF and D0 measurements for the Tevatron combination.}
\label{tbl:wgttev}
\end{table*}

\begin{table*}[!tbp]
  \centering
  \input{LHC_weight_table_precomb.tex}
  \caption{Relative weights (in percent) of the ATLAS and LHCb measurements for the LHC combination.}
\label{tbl:wgtlhc}
\end{table*}

\begin{table*}[!tbp]
  \centering
  \input{All_weight_table_precomb.tex}
  \caption{Relative weights (in percent) of individual measurements 
    for the combination of all available measurements.}
\label{tbl:wgtall}
\end{table*}

\begin{table*}[!tbp]
  \centering
  \input{All-LEP_weight_table_precomb.tex}
  \caption{Relative weights (in percent) of individual measurements 
    for the combination of all except the LEP measurement.}
\label{tbl:wgtallexclep}
\end{table*}

\begin{table*}[!tbp]
  \centering
  \input{All-CDF_weight_table_precomb.tex}
  \caption{Relative weights (in percent) of individual measurements 
    for the combination of all except the CDF measurement.}
\label{tbl:wgtallexccdf}
\end{table*}

\begin{table*}[!tbp]
  \centering
  \input{All-D0_weight_table_precomb.tex}
  \caption{Relative weights (in percent) of individual measurements 
    for the combination of all except the D0 measurement.}
\label{tbl:wgtallexcd0}
\end{table*}

\begin{table*}[!tbp]
  \centering
  \input{All-ATLAS_weight_table_precomb.tex}
  \caption{Relative weights (in percent) of individual measurements 
    for the combination of all except the ATLAS measurement.}
\label{tbl:wgtallexcatlas}
\end{table*}

\begin{table*}[!tbp]
  \centering
  \input{All-LHCb_weight_table_precomb.tex}
  \caption{Relative weights (in percent) of individual measurements 
    for the combination of all except the LHCb measurement.}
\label{tbl:wgtallexclhcb}
\end{table*}

%% file: TeVatron_weight_table_precomb.tex
\begin{tabular}{l@{\extracolsep{\fill}}rrrrrrrr}
  \hline \hline
          Measurement         & &   ABMP16  &  CT14  &    CT18  &  MMHT2014  &  MSHT20  &   NNPDF3.1  &   NNPDF4.0 \\
\hline 
                      CDF     & &   85.3  &    86.8  &    88.1  &    86.4  &    86.3  &    86.2  &    86.1 \\
                      D0      & &   14.7  &    13.2  &    11.9  &    13.6  &    13.7  &    13.8  &    13.9 \\
\hline \hline 
\end{tabular}

%% file: LHC_weight_table_precomb.tex
\begin{tabular}{l@{\extracolsep{\fill}}rrrrrrrr}
  \hline \hline
Measurement       &  &  ABMP16  &    CT14  &      CT18  &  MMHT2014  &    MSHT20  &   NNPDF3.1  &   NNPDF4.0 \\
\hline 
         ATLAS    &  &    77.8  &    69.3  &    70.5  &    72.4  &    73.7  &    74.9  &    77.0 \\
         LHCb     &  &    22.2  &    30.7  &    29.5  &    27.6  &    26.3  &    25.1  &    23.0 \\
\hline \hline
\end{tabular}

%% file: All_weight_table_precomb.tex
\begin{tabular}{l@{\extracolsep{\fill}}rrrrrrrr}
  \hline \hline
Measurement                 &     &    ABMP16  &      CT14  &      CT18  &  MMHT2014  &    MSHT20  &   NNPDF3.1  &   NNPDF4.0 \\
\hline
                      ATLAS &     &    19.9  &    28.2  &    28.2  &    20.2  &    22.8  &    19.7  &    24.8 \\
                      LHCb  &     &     6.0  &    12.7  &    13.4  &     9.7  &    10.7  &     8.5  &     8.9 \\
                      CDF   &  &    58.8  &    42.2  &    41.1  &    54.0  &    50.6  &    56.1  &    51.0 \\
                      D0    &  &    10.1  &     6.0  &     5.1  &     8.3  &     7.9  &     8.7  &     8.1 \\
                      LEP   &     &     5.2  &    10.9  &    12.2  &     7.7  &     8.0  &     7.0  &     7.2 \\
\hline \hline
\end{tabular}

%% file: All-LEP_weight_table_precomb.tex
\begin{tabular}{l@{\extracolsep{\fill}}rrrrrrrr}
  \hline \hline
Measurement                     &  &  ABMP16  &      CT14  &      CT18  &  MMHT2014  &    MSHT20  &   NNPDF3.1  &   NNPDF4.0 \\
\hline 
                          ATLAS  &    &    21.0  &    31.6  &    32.1  &    21.9  &    24.7  &    21.2  &    26.7 \\
                          LHCb   &    &     6.3  &    14.3  &    15.3  &    10.5  &    11.7  &     9.1  &     9.5 \\
                          CDF    & &    62.1  &    47.3  &    46.8  &    58.6  &    55.0  &    60.4  &    55.0 \\
                          D0     & &    10.6  &     6.7  &     5.8  &     9.0  &     8.6  &     9.3  &     8.7 \\
\hline \hline 
\end{tabular}

%% file: All-CDF_weight_table_precomb.tex
\begin{tabular}{l@{\extracolsep{\fill}}rrrrrrrr}
  \hline \hline
Measurement             &               &    ABMP16  &      CT14  &      CT18  &  MMHT2014  &    MSHT20  &   NNPDF3.1  &   NNPDF4.0 \\
\hline 
                           ATLAS   &   &    47.4  &    41.5  &    42.2  &    42.5  &    44.2  &    44.4  &    47.5 \\
                          LHCb  &     &    13.7  &    18.6  &    18.4  &    17.1  &    17.0  &    15.8  &    14.9 \\
                          D0 &    &    27.5  &    24.6  &    23.1  &    26.9  &    25.7  &    26.8  &    25.3 \\
                        LEP  &      &    11.5  &    15.3  &    16.2  &    13.5  &    13.1  &    13.0  &    12.4 \\
\hline \hline 
\end{tabular}

%% file: All-D0_weight_table_precomb.tex
\begin{tabular}{l@{\extracolsep{\fill}}rrrrrrrr}
  \hline \hline
Measurement        &                &    ABMP16  &      CT14  &      CT18  &  MMHT2014  &    MSHT20  &   NNPDF31  &   NNPDF40 \\
\hline
                           ATLAS &     &    21.9  &    28.8  &    28.7  &    21.5  &    24.0  &    21.3  &    26.2 \\
                              LHCb  &     &     6.5  &    13.0  &    13.6  &    10.1  &    11.1  &     8.9  &     9.2 \\
                         CDF  &  &    66.0  &    47.2  &    45.4  &    60.4  &    56.7  &    62.5  &    57.1 \\
                               LEP    &    &     5.6  &    11.1  &    12.3  &     8.0  &     8.2  &     7.3  &     7.5 \\
\hline \hline 
\end{tabular}

%% file: All-ATLAS_weight_table_precomb.tex
\begin{tabular}{l@{\extracolsep{\fill}}rrrrrrrr}
  \hline \hline
Measurement &               &    ABMP16  &      CT14  &      CT18  &  MMHT2014  &    MSHT20  &   NNPDF3.1  &   NNPDF4.0 \\
\hline
LHCb   &    &     7.2  &    14.8  &    15.7  &    10.4  &    12.1  &     9.4  &    11.0 \\
                CDF   &  &    73.8  &    61.0  &    60.4  &    69.5  &    67.4  &    71.0  &    68.7 \\
                         D0    &  &    12.7  &     9.2  &     7.9  &    10.9  &    10.7  &    11.3  &    11.1 \\
                                 LEP &       &     6.3  &    15.0  &    16.0  &     9.2  &     9.8  &     8.3  &     9.2 \\
\hline \hline 
\end{tabular}

%% file: All-LHCb_weight_table_precomb.tex
\begin{tabular}{l@{\extracolsep{\fill}}rrrrrrrr}
  \hline \hline
Measurement  &   &    ABMP16  &      CT14  &      CT18  &  MMHT2014  &    MSHT20  &   NNPDF3.1  &   NNPDF4.0 \\
\hline 
                   ATLAS  &    &    21.0  &    30.1  &    30.7  &    21.1  &    24.6  &    21.0  &    27.1 \\
                   CDF &      &    62.7  &    50.0  &    48.8  &    60.8  &    57.3  &    61.7  &    56.1 \\
                   D0   &   &    10.8  &     7.2  &     6.2  &     9.4  &     9.0  &     9.6  &     8.9 \\
                 LEP  &      &     5.5  &    12.7  &    14.2  &     8.7  &     9.1  &     7.7  &     8.0 \\
\hline \hline
\end{tabular}